\documentclass[fleqn,usenatbib]{mnras}

\usepackage{newtxtext,newtxmath}
\usepackage[usenames]{xcolor}
\usepackage{color}
\usepackage{tabularx}
\usepackage{comment}
\usepackage[normalem]{ulem}

\usepackage[T1]{fontenc}
\usepackage{ae,aecompl}

\usepackage{graphicx}	
\usepackage{amsmath}	

\usepackage{amssymb}	
\usepackage{bm,times}

\newcommand{\gd}[1]{\textcolor{purple}{\textbf{GD:} #1}}

\newcommand{\cp}[1]{\textcolor{red}{\textbf{CP:} #1}}

\title[Binary-disc dynamics in IRAS~04158+2805]{Circumbinary and circumstellar discs around the eccentric binary IRAS 04158+2805 --- a testbed for binary-disc interaction}

\author[E. Ragusa et al.]{\parbox{\textwidth}{
Enrico Ragusa$^{1}$\thanks{E-mail: er198@leicester.ac.uk}, Daniele Fasano$^{2}$, Claudia Toci$^{2}$, Gaspard Duch\^ene$^{3,4}$\\ Nicol\'as Cuello$^{4}$, Marion Villenave$^{5,4}$,  Gerrit van der Plas$^{4}$,
Giuseppe Lodato$^{2}$,\\ Fran\c{c}ois M\'enard$^{4}$, Daniel J. Price$^{6}$, Christophe Pinte$^{6,4}$, Karl Stapelfeldt$^{5}$, Schuyler Wolff$^{7}$}\vspace{0.15cm}
\\
$^{1}$School of Physics and Astronomy, University of Leicester, Leicester LE1 7RH, UK\\
$^{2}$Dipartimento di Fisica, Universit\'a degli Studi di Milano, via Celoria 16, 20133 Milano, Italy\\
$^{3}$Astronomy Department, University of California Berkeley, Berkeley, CA 94720, USA\\
$^{4}$Univ. Grenoble Alpes, CNRS, IPAG, F-38000 Grenoble, France.\\
$^{5}$Jet Propulsion Laboratory, California Institute of Technology, 4800 Oak Grove Drive, Pasadena, CA 91109, USA\\
$^{6}$School of Physics and Astronomy, Monash University, VIC 3800, Australia\\
$^{7}$Steward Observatory and the Department of Astronomy, The University of Arizona, 933 N Cherry Ave, Tucson, AZ, 85721, USA\\
}

\date{Accepted XXX. Received YYY; in original form ZZZ}

\pubyear{2021}
\begin{document}
\label{firstpage}
\pagerange{\pageref{firstpage}--\pageref{lastpage}}
\maketitle

\begin{abstract}
IRAS~04158+2805 has long been thought to be a very low mass T-Tauri star (VLMS) surrounded by a nearly edge-on, extremely large disc. Recent observations revealed that this source hosts a binary surrounded by an extended circumbinary disc with a central dust cavity. In this paper, we combine ALMA multi-wavelength observations of continuum and $^{12}$CO line emission, with H$\alpha$ imaging and Keck astrometric measures of the binary to develop a coherent dynamical model of this system.
The system features an azimuthal asymmetry detected at the western edge of the cavity in Band~7 observations and a wiggling outflow. Dust emission in ALMA Band 4 from the proximity of the individual stars suggests the presence of marginally resolved circumstellar discs. We estimate the binary orbital parameters from the measured arc of the orbit from Keck and ALMA astrometry. We further constrain these estimates using considerations from binary-disc interaction theory. We finally perform three SPH gas + dust simulations based on the theoretical constraints; we post-process the hydrodynamic output using radiative transfer Monte Carlo methods and directly compare the models with observations. Our results suggest that a highly eccentric $e\sim 0.5\textrm{--}0.7$ equal mass binary, with a semi-major axis of $\sim 55$ au, and small/moderate orbital plane vs. circumbinary disc inclination $\theta\lesssim 30^\circ$ provides a good match with observations. A dust mass of $\sim 1.5\times 10^{-4} {\rm M_\odot}$ best reproduces the flux in Band 7 continuum observations. Synthetic CO line emission maps qualitatively capture both the emission from the central region and the non-Keplerian nature of the gas motion in the binary proximity.
\end{abstract}




\section{Introduction}

The dynamics of accretion discs in the presence of a second mass orbiting the primary has been studied and discussed starting from the early 1980s, when an analytical expression for the tidal torque exerted by a satellite on the accretion disc in which it is embedded was provided \citep{lin1979,goldreich1980}. The original goal was to model how the moons of Saturn and Uranus perturbed the dynamics of their rings. Since then, the disc-satellite interaction framework has been improved both theoretically and numerically, discussing more generally the effects of the tidal interaction on the evolution of the properties of any type of binary objects (star+planets, star+star, black hole + black hole) and the discs surrounding them: the evolution of the binary semi-major axis (migration) and its orbital eccentricity (e.g. \citealp{syer1995,ward1997,ivanov1999,goldreich2003,ragusa2018,kanagawa2018}); the evolution of the disc structure, i.e. the formation of a gap or a cavity in the accretion disc (e.g. \citealp{artymowicz1994,goodman2001,pichardo2005}; \citealp{crida2006,pichardo2008,duffell2015,hirsh2020}); the modulation of accretion rate (e.g. \citealp{gunther2002,farris2014,dunhill2015,ragusa2016,munoz2018,teyssandier2020}) have all been discussed in the literature.

Despite these improvements on the theoretical and numerical side, direct observation of the effects of binaries on discs is recent. The advent of new generation interferometers, such as ALMA, and high contrast IR imaging, such as the SPHERE instrument mounted on the VLT, revealed a number of protoplanetary systems showing features such as: spirals, circular gaps, cavities, rings, horseshoes, clumps, shadows and warps \citep{garufi2017,andrews2018,avenhaus2018,vandermarel2018,long2018,garufi2020}.
The origin of these features has been mainly interpreted with the presence of planets or binary stars embedded in the accretion disc. Nevertheless, some alternative interpretations for individual structures exist, such as ice lines \citep{banzatti2015,zhang2015} for gaps, dead-zones inducing vortices for horseshoes \citep{regaly2012,regaly2017}, and multiple possibilities such as accretion from an external envelope \citep{harsono2011,lesur2015}, self-gravity roils \citep{lodato2004,cossins2010} or flybys for the formation of spirals \citep{clarke1993,Pfalzner2003,cuello2019}.

Discs surrounding resolved binary stars with high mass ratios ($q>0.1$) have also been detected. The system HD142527, that hosts a secondary M-dwarf \citep{biller12,lacour16,claudi2019}, is able to explain all the main features of the system \citep{price2018}. The system GG Tau A also hosts a large disc with a cavity carved by a hierarchical triple system -- that dynamically behaves as a binary -- with a mass ratio $q\sim 0.8\textrm{--}0.9$ \citep{cazzoletti2017,aly2018,keppler2020}. HD98800 shows polar alignment between the binary and the disc \citep{kennedy2019}. GW Ori, a hierarchical triple system, shows disc warping and tearing \citep{kraus2020}. Other recent detections of resolved binaries embedded in a circumbinary disc include: L1448 IRS 3B \citep{tobin2016}, L1551 IRS 5 \citep{cruzsaenz2019}, BHB 2007 11 \citep{alves2019}, IRAS 16293-2422 A \citep{maureira2020} and V892 Tau \citep{long2021}. In only a limited number of cases information about the binary dynamics is also available (e.g., \citealp{cazzoletti2017,price2018,long2021}).

In this paper, we discuss and interpret the multi-band observations of the system IRAS~04158+2805 \citep{glauser2008,andrews2008,villenave2020}.
This system was suggested to host an approximately equal mass binary star in its centre, as the central source showed inconsistencies between its spectral type and measured kinematic mass \citep{andrews2008}. Bright emission at radio wavelengths (ALMA, Band 4 continuum, 2.06 mm) from the circumstellar discs surrounding the two stars of the binary have been spatially resolved for the first time by \citet{villenave2020}, confirming that two gravitationally bound stars are in fact present at the centre of this system (projected separation of $\sim 0\farcs18$ i.e., $\sim 25$ au at $130$ pc). The system IRAS~04158+2805 interestingly features a circumbinary disc with a cavity of $\sim 200$ au in radius and a prominent azimuthal over-density located at the western edge of the cavity.

We complete the available information about the structure of the system --- provided by the multi-wavelength observations by \citet{villenave2020} (ALMA Band 4, 2.06~mm, and Band 7, 0.890~mm) --- by presenting: a) CO J=3-2 line emission and high resolution ALMA Band 7 continuum  observations (acquired during the \citealt{villenave2020} survey, but not previously published); b)  measures of the astrometric motion of the binary at different epochs using ALMA and Keck adaptive optics between 2013 and 2019, allowing us to gain insights about the binary orbital parameters.

We also present optical observations performed using the Canada France Hawaii Telescope 12K camera (CFHT12K) with an H-$\alpha$ filter, that shows the presence of a wiggling jet, launched from the centre of the system and possibly related to the presence of the binary.

Due to the availability of data at multiple wavelengths, complexity of the structures and information about the binary orbital motion, this system represents a unique laboratory to test binary-disc interaction theory through a direct comparison with observations.

The paper is divided as follows: in Sec. \ref{sec:obs} we present the observations of IRAS~04158+2805 that we will use for our analysis. In Sec. \ref{sec:obsdiscussion} we discuss the morphology and available observational information about the system. In Sec. \ref{sec:anest} we infer the orbital properties of the binary using the astrometric data and we further constrain them in the light of the system structure using analytical considerations; we also provide quantitative estimates of the amount of gas and dust present in the system from the observed luminosity. In Sec. \ref{sec:SPH} we introduce the numerical gas+dust and Monte Carlo radiative transfer simulations we performed in order to provide a direct comparison of our predictions with observations. In Sec. \ref{sec:discussion} we discuss the results and we finally draw our conclusions in Sec. \ref{sec:conclusions}.

\section{Observations of IRAS~04158+2805}\label{sec:obs}

\subsection{Near-infrared imaging}

\begin{figure}
    \centering
    \includegraphics[width=\columnwidth]{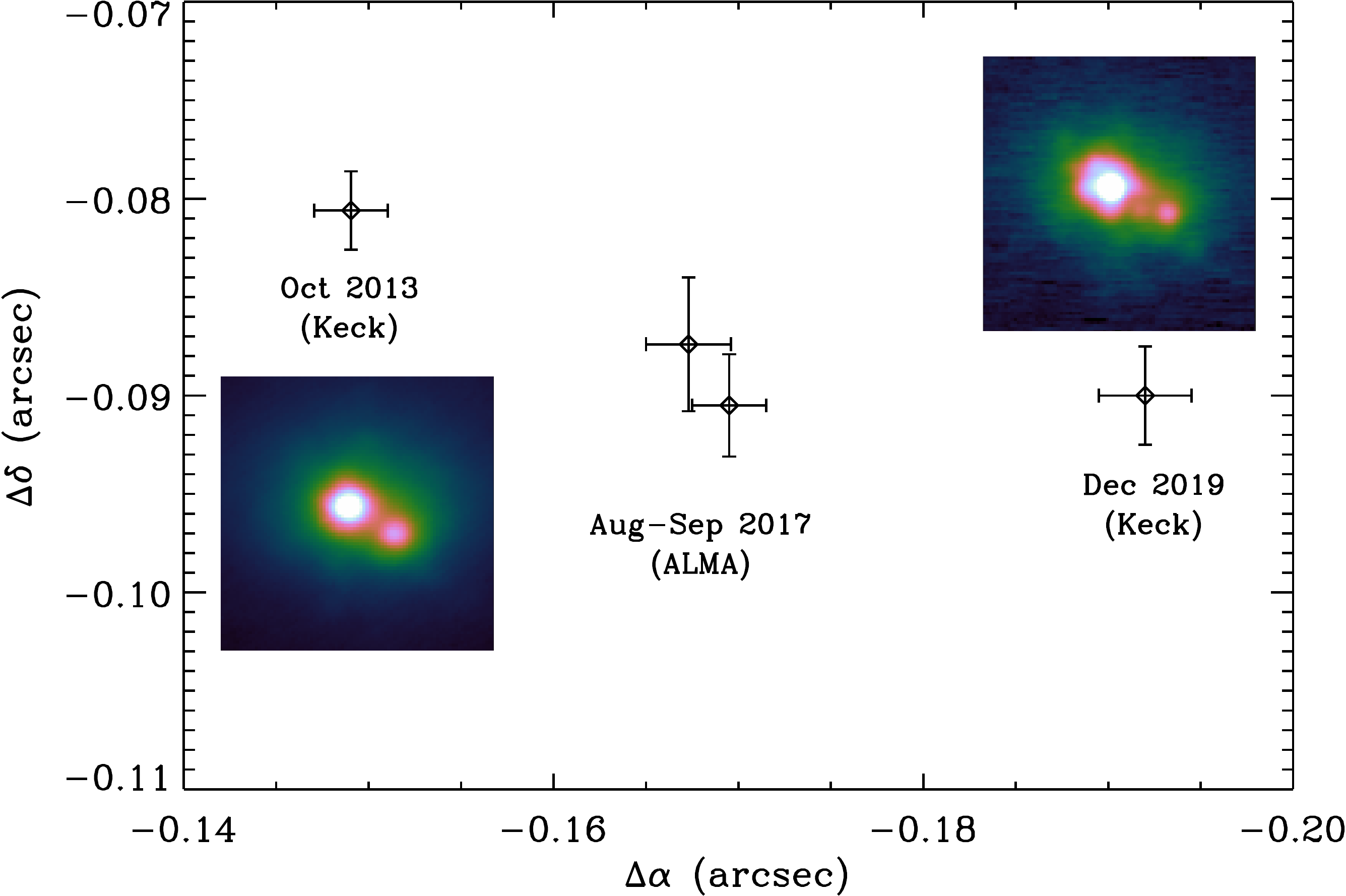}
    \caption{Astrometric motion of the binary along with two insets showing the 2013 and 2019 Keck AO images.
    }
    \label{fig:keck}
\end{figure}

On Dec 20, 2019, we obtained laser guide star adaptive optics imaging of IRAS~04158+2805 using NIRC2 on the Keck\,II telescope. We used the narrow camera ($\approx0\farcs01$/pix) and the $K'$ filter (2.12\,$\mu$m) and obtained 15 short, unsaturated images to resolve the close binary. Coupled with co-adding and on-chip dithering to estimate the sky background, we obtained a total on-source integration time of 150\,s. Data reduction consisted of the usual steps of sky subtraction, flat field, image registration and median combination. A few frames were removed from the stack as they suffered from significantly inferior image quality. While conditions were good overall, the faint and somewhat distant tip-tilt guide star used in the observations (R$\approx$17, 22\arcsec\ away) resulted in a modest Strehl ratio. The FWHM of the point sources in the final image was about 0\farcs09, still sufficient to resolve the binary (see Fig. \ref{fig:keck}). The relative astrometry of the binary was estimated by performing Gaussian fitting to the two components. The binary separation was 0\farcs212$\pm$0\farcs003 and its position angle, measured East from North, 244\fdg9$\pm$0\fdg7. The flux ratio between the component is about 2.2$\pm0.1$\,mag.

We also analyzed archival NIRC2 data taken on Oct 24 2013 with a similar set-up as described above (PI: T. Dupuy). Images using the $J$, $H$, $K'$ and $L'$ (1.25, 1.63, 2.12 and 3.78\,$\mu$m, respectively) were obtained with total on-source integration times ranging from 120 to 240\,s. Data reduction followed the same process as described above. The FWHM of point sources ranged from 0\farcs09 to 0\farcs11, with the highest (lowest) resolution achieved at $K'$ ($L'$). The binary was well resolved at all wavelengths (see Fig. \ref{fig:keck}) and its relative astrometry was again estimated using Gaussian fits, yielding a separation of 0\farcs169$\pm$0\farcs002 and a position angle of 241\fdg6$\pm$0\fdg7 (after averaging over all wavelengths). We estimated flux ratios of 1.8, 1.6, 1.3 and 1.1\,mag from $J$ to $L'$, with a typical uncertainty of 0.1\,mag.

\subsection{H$\alpha$ imaging}

\begin{figure}
    \centering
    \includegraphics[width=\columnwidth]{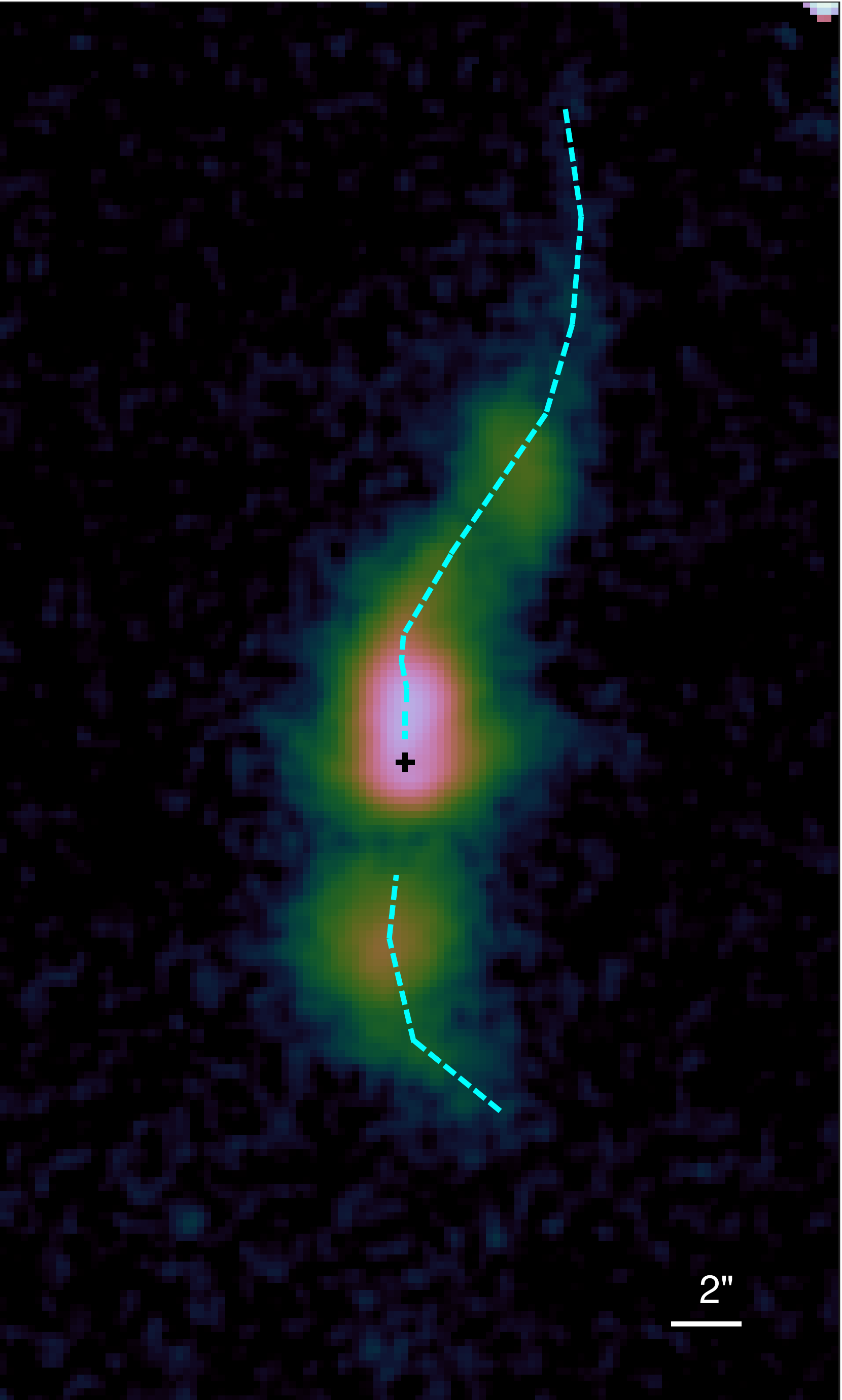}
    \caption{Continuum-subtracted CFHT12K H-$\alpha$ image of IRAS~04158+2805, using the $I$ band image as a jet-free estimate of the continuum. North is up, East to the left and the field of view of 25\arcsec$\times40$\arcsec. The black cross marks the brightest point in the continuum image, close to but likely slightly offset from the location of the central stars due to scattering off the disc. The cyan curve is a guide to the morphology of the H$\alpha$ jet.}
    \label{fig:hstwiggle}
\end{figure}

IRAS~04158+2805 was observed on Dec 29 1999 with CFHT12K, a wide-field $\approx$0\farcs2/pix visible camera installed on the Canada-France-Hawaii Telescope (CFHT), with the broadband $R$ and $I$ filters (650 and 820\,nm, respectively) and with the narrowband H$\alpha$ filter (centered at 658\,nm and with a 7\,nm bandwidth). Total integration times were 2$\times$600 s, 3$\times$1200 s and 2$\times$300 s with the $R$, H$\alpha$ and $I$ filters, respectively. Observational conditions were worse than average, with point sources having measured FWHM of 1\farcs0--1\farcs1. The data reduction was performed within the {\sc elixir} environment, a data processing pipeline developed at CFHT for intensive large scale format data reduction. The process includes the standard steps of bias subtraction, flatfield correction and fringe removal (the latter only applies to the $I$ image). The resulting frames were then registered and averaged to form the final images (Fig. \ref{fig:hstwiggle}).

In this study, we are particularly interested in the H$\alpha$ image, which reveals a jet launched by the central binary system. Even in this narrow-band filter, the image contains a contribution from photon scattered off the disc surface, so we decided to subtract a continuum image to highlight the jet emission. Counter-intuitively, the $R$ filter is not ideal for this purpose as H$\alpha$ emission from the jet contaminates that image, which would lead to self-subtraction of the jet. We therefore opted to use the $I$ image of the disc, in which no evidence of the jet is found \citep[see also][]{glauser2008}. We scaled down the $I$ image and subtracted it from the H$\alpha$ image, aiming at obtaining zero signal in regions located far from the jet axis. Since the exact disc morphology could be different between $R$ and $I$, this process is likely imperfect close to the disc itself but should have no consequences for most of the jet emission, which is found far above/below the disc.

\subsection{ALMA observations}
IRAS~04158+2805 was observed with ALMA as part of a survey of 12 highly inclined discs (Project 2016.1.00460.S, PI: M\'enard; \citealt{villenave2020}).

In this work, we combine previously published continuum ALMA Band 4 (2.06 mm) and Band 7 (0.89 mm, compact configuration) observations of the source (we refer to section 2.2 of \citealt{villenave2020} for details about the data reduction) with previously unpublished $^{12}$CO line emission and higher resolution Band 7 continuum observations from the same program.

\begin{figure*}
	\includegraphics[trim={1.7cm 2cm 2.2cm 3cm},clip,width=0.98\textwidth]{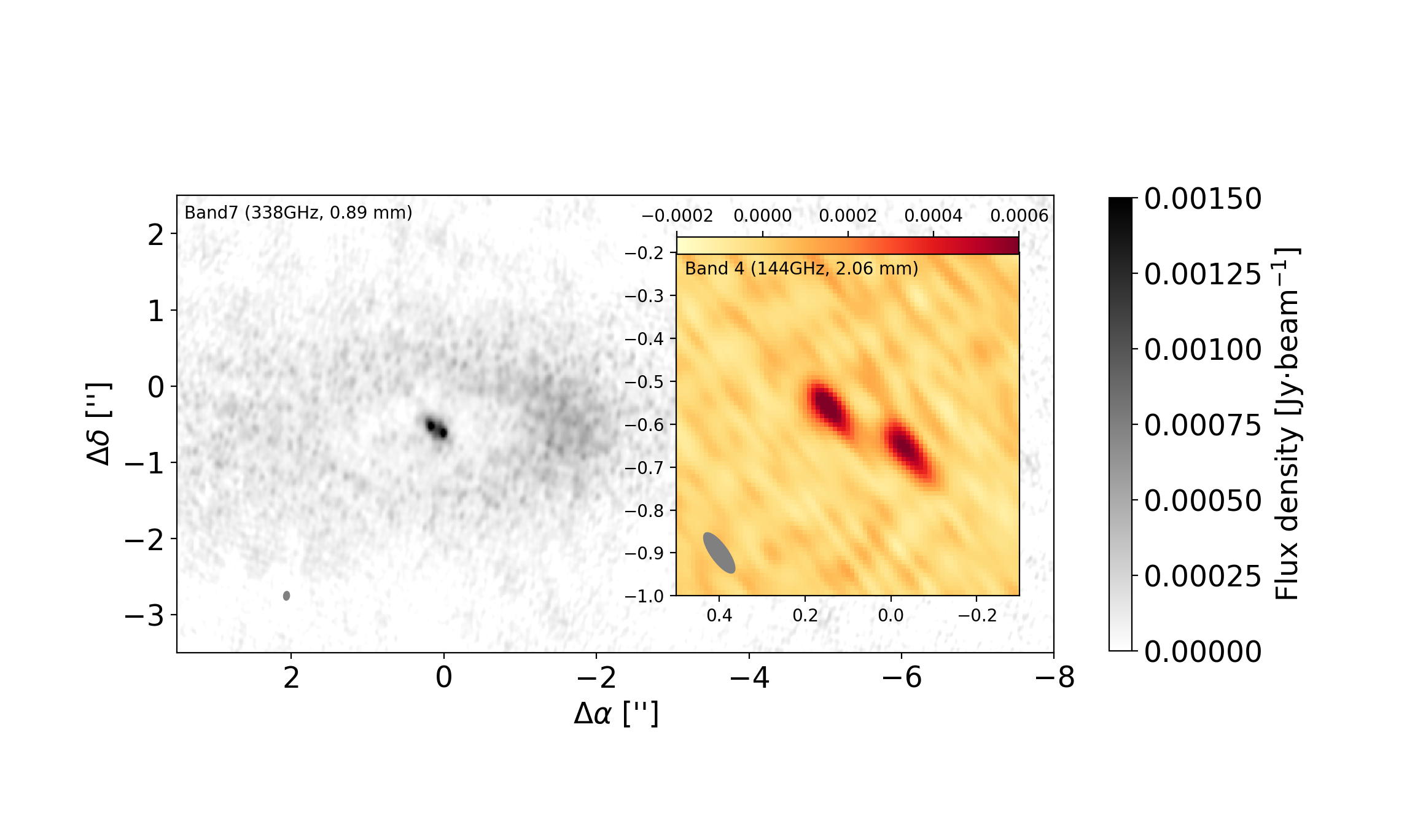}
\caption{ALMA observations of IRAS~04158+2805, Band 7 continuum compact + extended configuration (grey plot, from \citealp{villenave2020} survey but previously unpublished) and zoom-in on the circumstellar discs in Band 4 continuum from \citet{villenave2020}. The (0,0) pointing centre for this image is at RA (h m s): 04 18 58.1, Dec ($^\circ\, '\, ''$): +28 12 23.4. The declination of the pointing centre is off-centred by $\sim 0\farcs6$.
}\label{fig:contB4B7obs}
\end{figure*}

The Band 4 observations (previously published in \citealp{villenave2020}) were performed on 27th of September 2017 with four continuum spectral windows centered on 138, 140, 150, and 152 GHz. We produce a continuum image from the calibrated visibilities using the \texttt{clean} function in CASA, with a Briggs robust parameter of 0.5, achieving an angular resolution of 0\farcs093$\times$0\farcs038 (zoom-in window of Fig. \ref{fig:contB4B7obs}). The rms noise for this configuration is $0.04$ mJy/beam.

\begin{figure*}
    \centering
    \includegraphics[trim={2.cm 0cm 2.9cm 1cm},clip,width=\columnwidth]{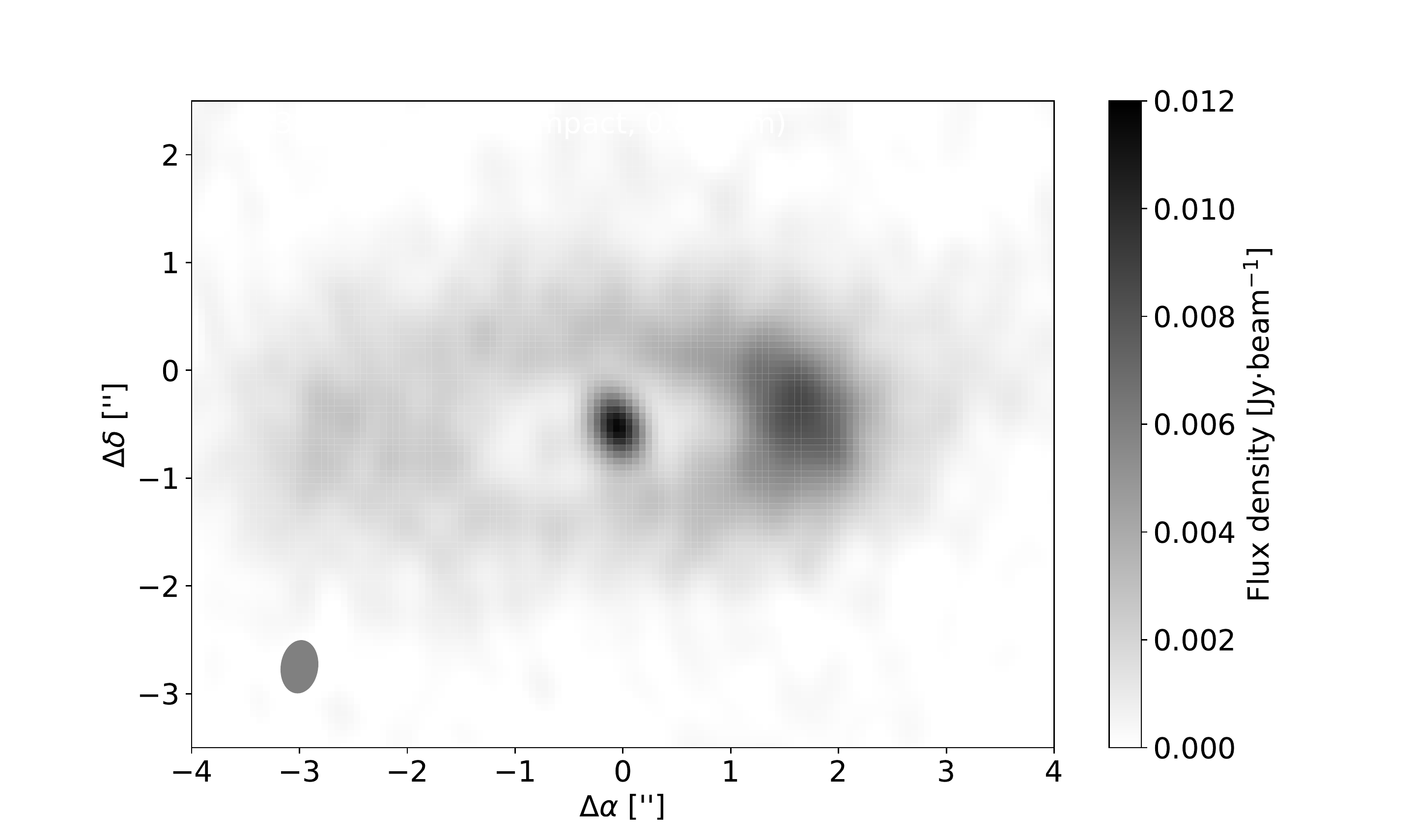}
    \includegraphics[trim={2.3cm 0cm 2.7cm 1cm},clip,width=\columnwidth]{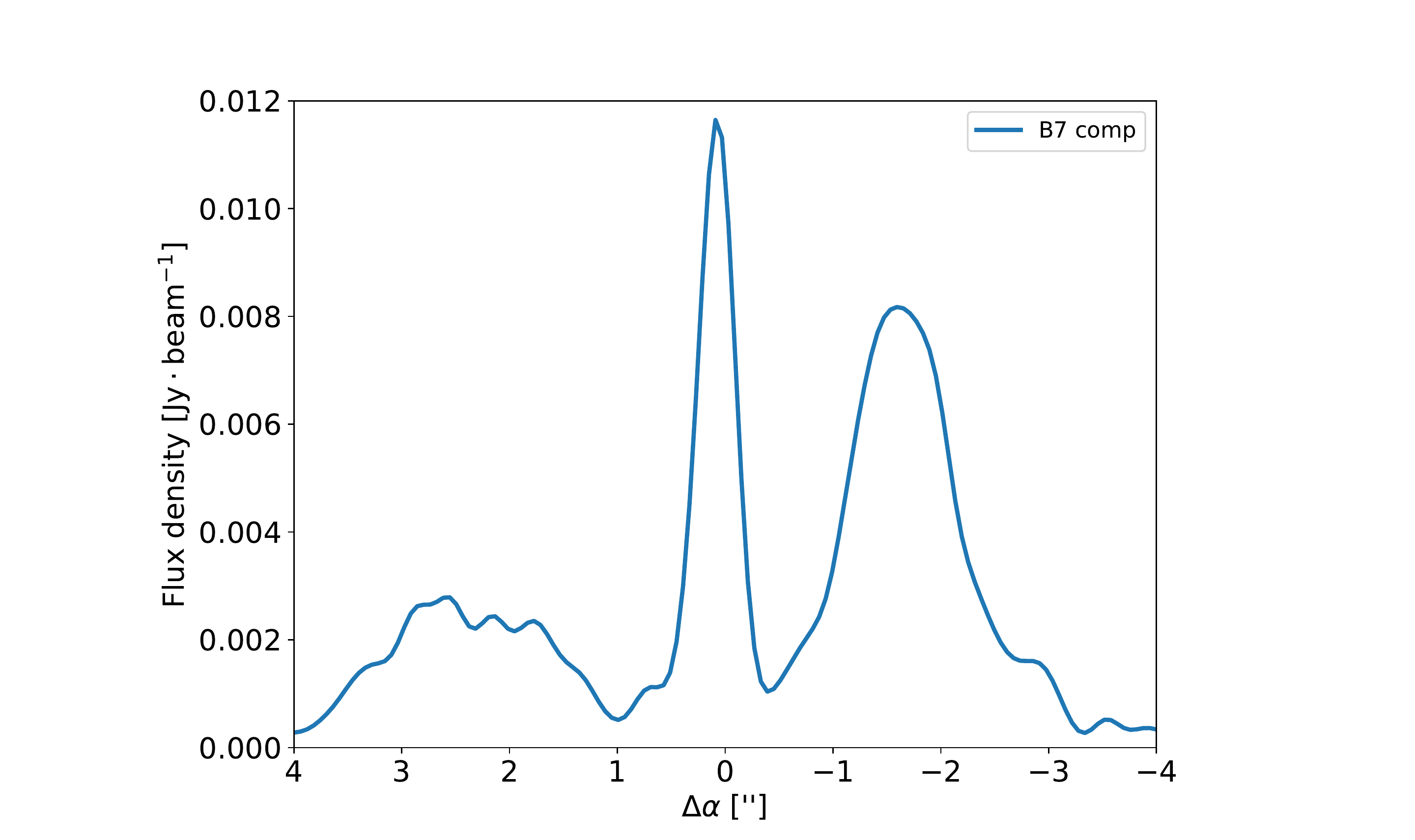}
    \caption{Left panel: ALMA observations of IRAS 04158+2805, Band 7 continuum using only the compact configuration.  Right panel: cut along the major-axis of the cavity from image in left panel.
    }
    \label{fig:cutsObs}
\end{figure*}

The Band 7 observations were performed with two array configurations: a compact configuration, observed on the 24th of November 2016 with maximum baseline of 0.7 km (previously published in \citealp{villenave2020}), and an extended configuration, observed on August 18, 2017 with a maximum baseline of 3.6 km (previously unpublished). The spectral set-up included three continuum spectral windows, centered on 344.5, 334.0, and 332.0 GHz, and one spectral window set to observe the $^{12}$CO J=3-2 transition at 345.796GHz. We produce two continuum images from the visibilities, using the CASA task \texttt{clean}, with a robust Briggs parameter of 0.5. The first image combines the compact and extended configuration observations and achieves an angular resolution of 0\farcs108$\times$0\farcs070 (main panel of Fig. \ref{fig:contB4B7obs}), rms noise is $0.07$ mJy/beam for band 7 compact + extended configuration. The second image is extracted from the compact array observations only and achieves an angular resolution of 0\farcs481$\times$0\farcs332 (left panel of Fig. \ref{fig:cutsObs}), rms noise for this configuration is $0.2$ mJy/beam.

In both the extended array Band 7 and the Band 4 observations, the central binary is resolved in two point-like components that represent the compact circumstellar discs (Fig. \ref{fig:contB4B7obs}, see also \citealp{villenave2020}). We derived the relative position of the binary from Gaussian fits to the two continuum maps, resulting in separations of 0\farcs188$\pm$0\farcs003 and 0\farcs192$\pm$0\farcs002 and position angles of 242\fdg4$\pm$1\fdg0 and 241\fdg9$\pm$0\fdg7 in Band 7 and Band 4, respectively.

\begin{figure*}
    \includegraphics[trim={4.cm 0cm 2.7cm 1cm},clip,width=0.49\textwidth]{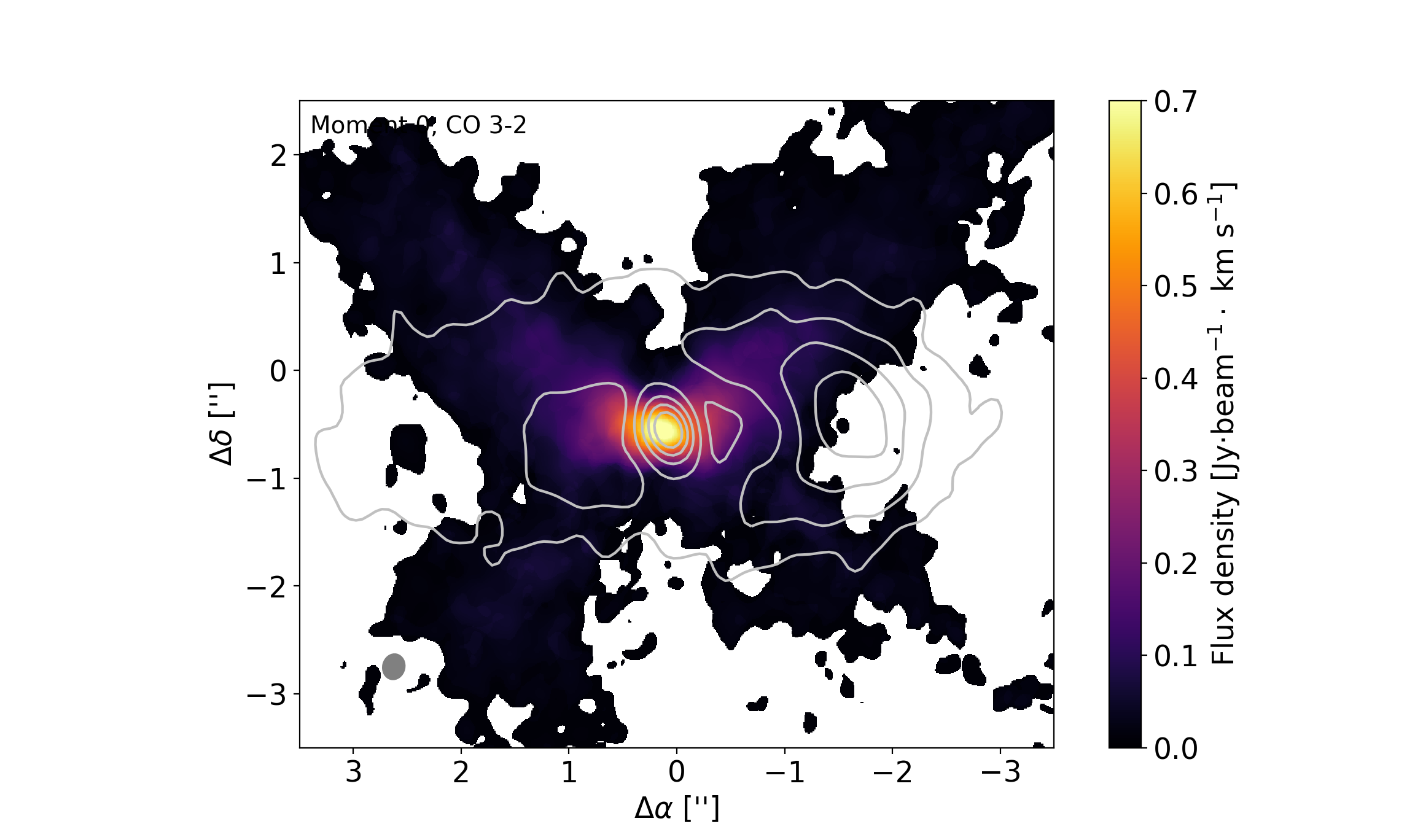}
	\includegraphics[trim={4.cm 0cm 2.7cm 1cm},clip,width=0.49\textwidth]{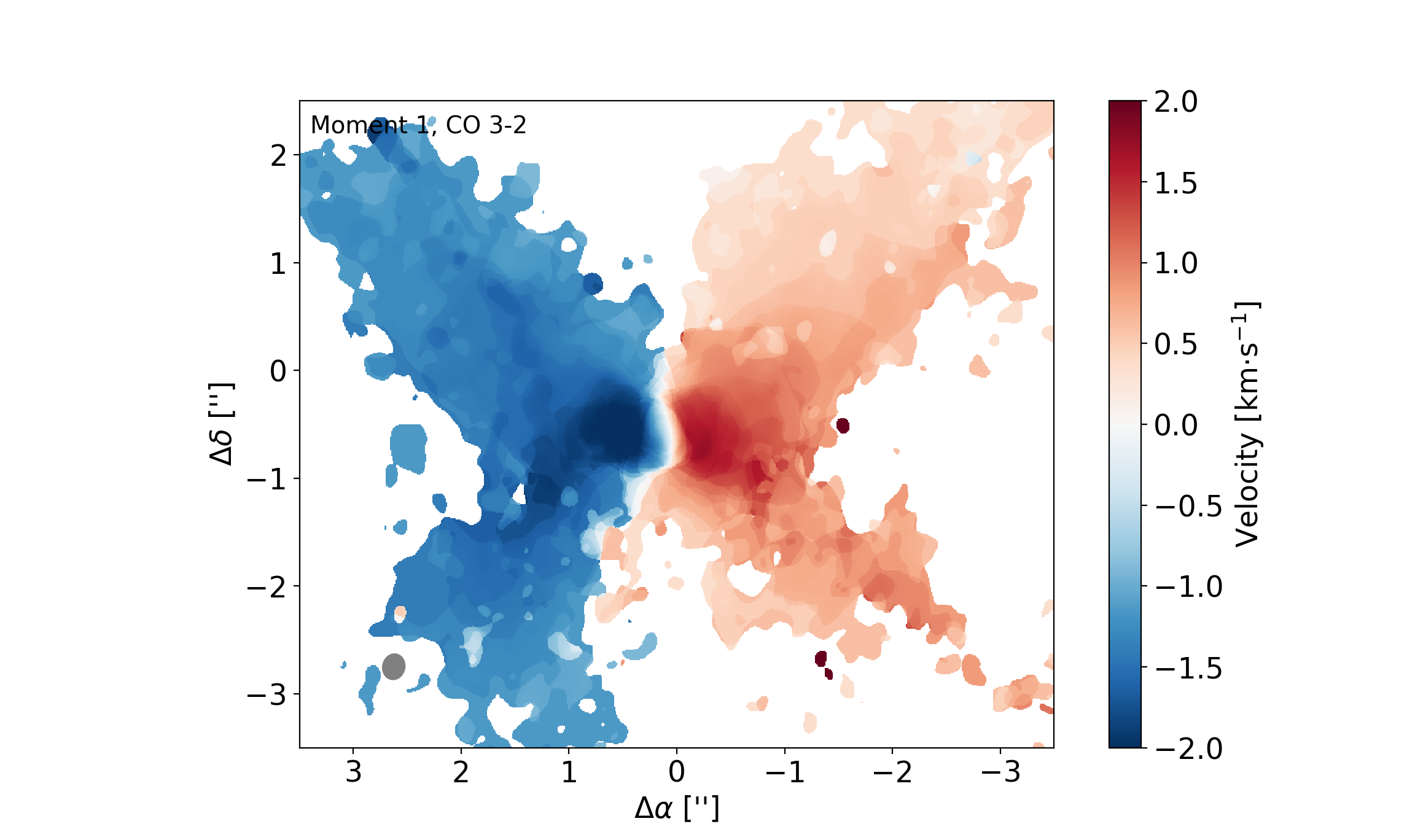}
	\caption{Observed moment 0 and moment 1 maps of CO $J=3-2$ line emission (left and right panel, respectively). Contours superimposed to the moment 0 map show emission levels of flux density $I_\nu=\{1.5;3.;5.;7.;9.\}$ mJy/beam from the compact array configuration data only, for reference. Moment 1 maps were obtained subtracting a constant velocity field of $7.7 \,{\rm km\,s^{-1}}$ to account for the proper motion of the system.  }\label{fig:MmapsObs}
\end{figure*}

Additionally, we present $^{12}$CO J=3--2 observations of IRAS~04158+2805. After substracting the continuum from the line emission with the \texttt{uvcontsub} CASA task, we extracted the $^{12}$CO channel maps from the visibilities of both the compact and extended configurations, using the \texttt{tclean} task. We used a velocity resolution of 0.25 km/s, the multiscale option and a Briggs robust parameter of 0.5. Additionally, to increase the signal to noise, we used a uvtaper. We achieve an angular resolution of 0\farcs235$\times$0\farcs201. Finally, we extracted the moment 0 (integrated line emission) and moment 1 maps (velocity gradient) using the \texttt{immoments} function of CASA. The moment maps were generated over pixels brighter than 3 times the rms and included within the mask used to clean the different channel maps. We show the moment maps in Fig. \ref{fig:MmapsObs}.

\section{IRAS~04158+2805: a circumbinary disc around a resolved binary}\label{sec:obsdiscussion}

The system IRAS~04158+2805 is a young source located in the Taurus star forming region.

The near- to mid-infrared spectral index of the system is positive, classifying it in the Class I category (e.g., \citealp{white2004}). In general, this is interpreted as a star + disc system embedded in a remnant envelope that is still opaque at near-infrared wavelengths. However, at high inclinations, this interpretation can be incorrect as the disc itself can block the starlight even in the absence of an envelope, thus mimicking a younger, embedded system (e.g., \citealp{whitney2003}, \citealp{crapsi2008}). The interpretation of the spectral index is thus ambiguous and additional envelope tracers are necessary to assert the true nature of the system (e.g., \citealp{vankempen2009}). In the case of IRAS~04158+2805, such tracers are not available. \citet{sheehan2017} reproduced the SED of the system and low-resolution submillimeter observations with a disc + envelope (Class I) model, while \citet{glauser2008} reproduced the SED and scattered light images with a simpler disc model. While we cannot definitively exclude the presence of an envelope in the system, we note that the large disk radius could account for the large scale submillimeter emission, and thus adopt a simpler envelope-free model in line with \citet{glauser2008} and \citet{park2002}.

Parallactic measures of its distance provided contrasting results: 89.7$\pm$4.6\,pc in Gaia DR2, 128.9$\pm$8.9\,pc in EDR3, but in either case there are highly significant residuals (Renormalised Unit Weight Error -- goodness-of-fit statistic, ${\rm RUWE}\sim 1.7$) indicating a poor astrometric fit of the Gaia data -- from EDR3 documentation ${\rm RUWE}<1$ indicates a reliable estimate of the parallax. Given the nebulous nature of the source in the optical, it is reasonable to expect some/most of the parallactic signal to be confused with brightness changes. For this reason, we will use an average distance of the sources in the same part of the Taurus cloud: $D=130.3\pm0.6$\,pc in the B10 cloud (group 7 in \citealt{galli2019}) and $D=130.5\pm0.5$\,pc in L1495 (Taurus A group in \citealt{roccatagliata2020}).

In this section we discuss the properties of the binary and discs from current and past observations, that constitute the basis for a more quantitative analysis that we will provide in Sec. \ref{sec:anest}. A summary of the relevant properties discussed in Sec. \ref{sec:stars} and \ref{sec:IRASdiscs} is provided in Tab. \ref{tab:flux}.

\subsection{Stars/Binary in IRAS~04158+2805}\label{sec:stars}

As previously mentioned, the binary nature of IRAS~04158+2805 had already been speculated in the past.
Estimates of the total central mass in this system, based on pre-main sequence stellar models for its spectral type M5-6, predicted a mass $M_{\rm \star,tot}\sim 0.09\textrm{--}0.16\, {\rm M_\odot}$ (\citealt{white2004}, see also \citealt{sheehan2017} for a review of past observations).
Nevertheless, a kinematic measure of the total central mass in this system from CO 3-2 line emission constrained the central mass in the system to range $M_{\rm \star,tot}\sim 0.15\textrm{--}0.45 {\rm M_\odot}$ \citep{andrews2008}.

\begin{figure}
    \centering
    \includegraphics[trim={1.8cm 0.6cm 2cm 1.1cm},width=\columnwidth]{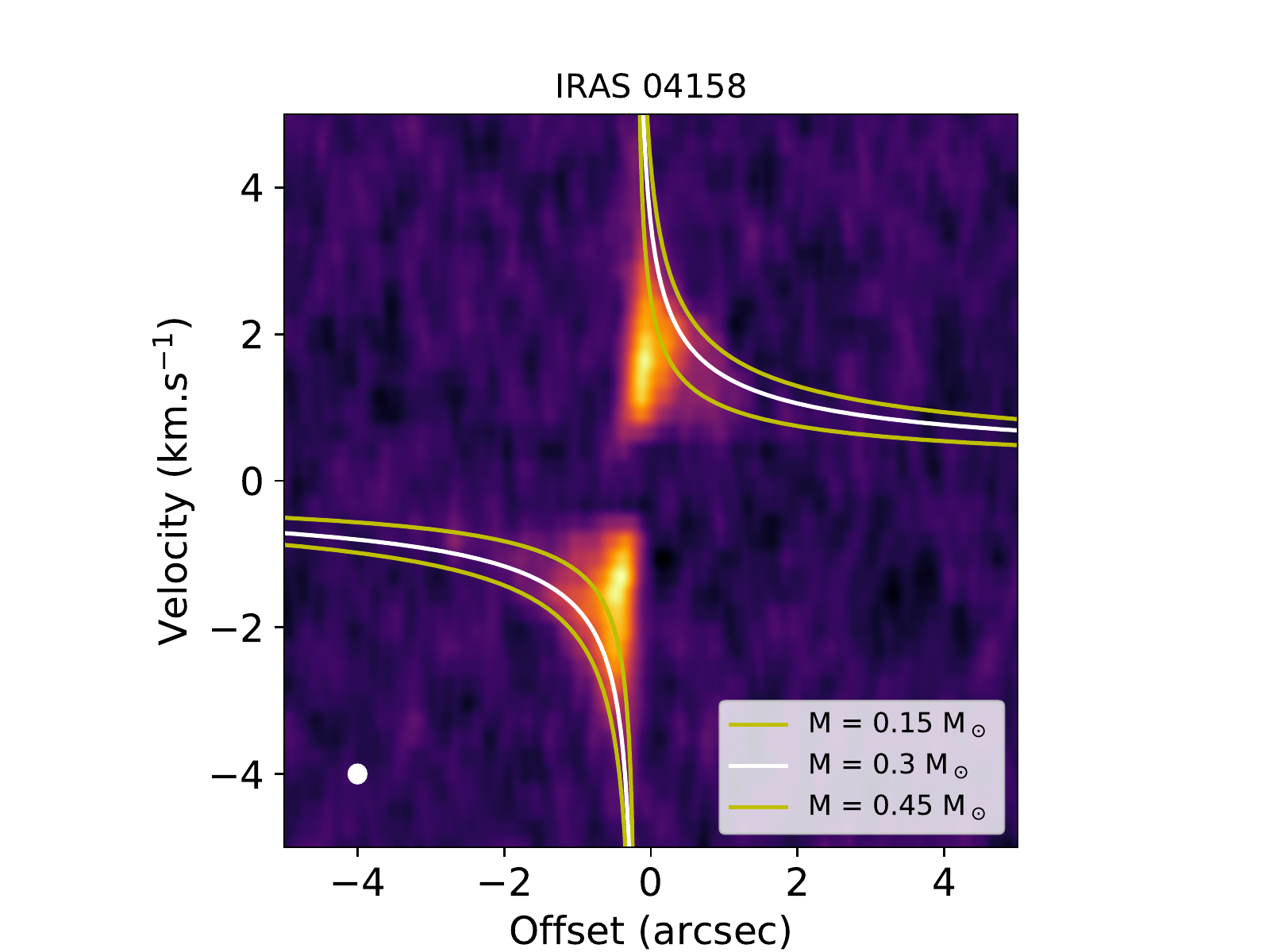}
    \caption{P--V diagram along the major-axis from ALMA CO line data. Velocity curves for single central masses $M_{\star,{\rm tot}}=\{0.15,0.3,0.45\}\, {\rm M_\odot}$, assuming an inclination of the disc $i_{\rm disc}=62^\circ$, and purely Keplerian motion are plotted in white and yellow. The ellipse in the left bottom corner shows spatial (major-axis of the beam) and spectral resolution of the diagram.}
    \label{fig:pvdiag}
\end{figure}

We performed an independent estimate using the same approach creating a position-velocity diagram --- from now on, P-V diagram (shown in Fig. \ref{fig:pvdiag}) --- with the ALMA CO line data and compared it with the predictions for different central masses -- predictions assume an inclination of $i_{\rm disc}=62^\circ$, a single central mass and purely Keplerian motion of the gas. The P-V diagram does not allow us to improve on the kinematic mass estimate by \citet{andrews2008}, and appears to confirm the same range of masses previously suggested.

Due to the previous inconsistency between the spectral type and kinematic mass estimates, \citet{andrews2008} speculated about the possible binary nature of this source.
\citet{villenave2020} confirmed the previous expectation that the central source in IRAS~04158+2805 is indeed a binary. Given the binary nature of the source, we note that any mass estimate based on the assumption of Keplerian motion of the material in the proximity of the binary should be treated as indicative and not as a precise measure. We will discuss in Sec. \ref{sec:discussion} how a change in the central mass might affect our analysis.

Both Keck (Fig. \ref{fig:keck}) and ALMA Band 4 and 7 (Fig. \ref{fig:contB4B7obs}) confirm the prediction of \citet{andrews2008} by showing the presence of resolved stellar and circumstellar emission at IR and radio wavelengths, respectively -- the latter from the dusty discs around the individual stars of the binary (as will be better discussed in the next Section). Such a binary has a projected separation $\sim 0\farcs18$, i.e. $\sim 25$ au at the source distance.

The spectral type mass estimates for the individual stars coupled with the kinematic total mass suggest the binary being approximately equal mass ($q=M_2/M_1\approx 1$).
However, the near-infrared images presented here show that one component is brighter than the other, by 1--2\,mag with both wavelength and temporal variations. Inferring a binary mass ratio from these flux ratios is a complicated exercise due to 1) the possible contamination by disc emission, and 2) the fact that absorption and scattering by the circumbinary disc affects (reduces) the apparent brightness of both components.
For simplicity, in the following sections the analysis will consider an equal mass binary, since a small change in the binary mass ratio would not affect significantly our results in any case.

\subsection{Discs in IRAS~04158+2805}\label{sec:IRASdiscs}

\begin{table*}
\begin{minipage}[t]{\textwidth}
\centering
\noindent\begin{tabularx}{\columnwidth}{@{\extracolsep{\stretch{1}}}{l}*{8}{c}@{}}
     \hline\hline
     Main system properties &$D$& 
     $F_{{\rm B7, tot}}$&  $F_{{\rm B7},i}$& 
     $F_{{\rm B4},i}$& 
     $R_{\rm cav}$& 
     $R_{{\rm cs},i}$ & $i_{\rm disc}$ & $\Omega_{\rm disc}$
     \\
     & [pc] & [mJy] & [mJy] & [mJy] & [au] & [au] & [$^\circ$] & [$^\circ$]\\
     \hline
     &$130.4\pm 0.6$  &
     $259\pm 26$  &
     $6.5\pm 0.8$  &
     $1\pm 0.1$  &
     $185$  &
     $3\textrm{--}7$  & $62\pm0.6$ &$93\pm 1$\\
     \hline\hline
\end{tabularx}
\end{minipage}
\caption{
Summary table of the main observational properties of the discs in IRAS~04158+2805: $D$ is distance from source \citep{galli2019,roccatagliata2020}; $F_{B7,{\rm tot}}$ is the total flux in ALMA band 7; $F_{{\rm B7},i}$ and $F_{{\rm B4},i}$ are the averaged (${\rm avg}[F_{B\star,1},F_{B\star,2}]$) emission from each circum-stellar disc in ALMA band 7 and band 4, respectively; $R_{\rm cav}$ is the cavity radius; $R_{{\rm cs},i}$ is the circum-stellar discs size; $i_{\rm disc}$ \citep{glauser2008,villenave2020} and $\Omega_{\rm disc}$ \citet{andrews2008}. \label{tab:flux}}
\end{table*}

Band 4 and Band 7 dust continuum ALMA images of IRAS~04158+2805 highlight the presence of three disc-like structures: one circumbinary disc with a large cavity and a crescent shaped over-density at its western edge, and two circumstellar discs surrounding the individual stars of the binary.

The extended circumbinary disc (total flux in Band 7 $F_{{\rm B7,tot}}=259\pm 26$\, mJy) is visible in Fig. \ref{fig:contB4B7obs}. The two circumstellar discs are marginally resolved in Band 4 and in the Band 7 compact + extended array image (see Fig. \ref{fig:contB4B7obs}) and are characterised by very similar mm fluxes --- each disc has a total flux of $F_{{\rm B4},i}=1\pm 0.1$\, mJy in Band 4 and $F_{{\rm B7},i}=6.5\pm 0.8$\, mJy in Band 7 (spectral index $\alpha_{B7-B4}\approx 2.25$) --- confirming the binary nature of the source at the centre of the system.

The cavity appears to be lopsided, with the apocentre located on the eastern side and pericentre located on the western one. It presents a bright azimuthal asymmetry in the dust emission on the western side (contrast-ratio West vs. East $\delta_\rho\sim 4$). With reference to the right panel of Fig. \ref{fig:cutsObs}, we measure the location where the brightness reaches 50\% of the peak value of the corresponding lobe and define the apocentre radius $R_{\rm apo}\sim 1\farcs65$, i.e. $R_{\rm apo}\sim 215$ au at the source distance, and pericentre radius $R_{\rm peri}\sim 1\farcs15$, i.e. $R_{\rm peri}\sim 150$ au. Note that the emission at the apocentre does not reach a clear peak, however the cavity edge is visible in the left panel of Fig. \ref{fig:cutsObs}. Such radii suggest a dust cavity semi-major axis $a_{\rm cav}\sim 185$ au and a cavity apparent eccentricity $e_{\rm cav}\sim 0.2$. Unless differently specified, we will refer to $a_{\rm cav}\sim 185$ au as the ``size'' of the cavity.

The size of the circumstellar discs slightly exceeds the beam along the minor axis in Band 4. We note that, given the geometry of the system, there is no reason to believe that the circumstellar discs are face-on: indeed, they are most likely to be inclined with respect to the plane of the sky. The projection effect might significantly reduce their apparent size and be masked by the elongated beam shape if they have a similar orientation. Considering the minor axis of the beam in Band 4 ($\sim 0\farcs04$, i.e. $\sim 5$ au), we believe that $R_{{\rm low}}\approx3$ au constitutes a reasonable lower limit for their size; whereas the beam major-axis $R_{{\rm up}}=7$ au constitutes the upper limit.

The disc aspect ratio has been inferred to be $H/R\sim 0.12\textrm{--} 0.2$ at $100$ au \citep{glauser2008,sheehan2017}. The circumbinary disc inclination has been estimated to be $i_{\rm disc}\sim 62\pm 0.6^\circ$  \citep{glauser2008, villenave2020}. The longitude of the ascending node\footnote{Note that the angle is calculated as the position angle, i.e. starting from the North axis and proceeding counter-clockwise in the plane of the sky.} of the disc is $\Omega_{\rm disc}\sim 93\pm 1^\circ$ (\citealp{andrews2008,sheehan2017}, i.e., the disc near side is South) and the material rotates counter-clockwise as can be noted in the moment 1 map (right panel of Fig. \ref{fig:MmapsObs}).

The moment 0 map of the CO gas line emission is brightest in the cavity area, highlighting the presence of gas up to the very central region of the system (Fig. \ref{fig:MmapsObs}). The emission is characterised by an X-shaped pattern which is brighter towards North --- consistent with the near side being South. The moment 1 maps show signs of non-Keplerianity in the cavity area, where the locus of points with projected velocity along the line of sight $v_{\rm proj}=0$ produces a S-shaped pattern \citep{rosenfeld2014}, which might indicate the presence of gas radial motion of tidal streams in the dust cavity region. We finally note that if also the gas is characterised by the same eccentricity as the dust ($e_{\rm cav}\sim 0.2$, previously obtained from geometrical considerations) we expect the velocity ratio between apocentre and pericentre to be \citep{murray1999}
\begin{equation}
    \frac{v_{\rm peri}}{v_{\rm apo}}=\frac{1+e_{\rm cav}}{1-e_{\rm cav}}\sim 1.5.
\end{equation}
However, such a velocity asymmetry is not evident neither from the moment 1 map in Fig. \ref{fig:MmapsObs} nor from the P-V diagram in Fig. \ref{fig:pvdiag}, implying that the gaseous disc is probably not as eccentric.

\section{Constraining the properties of the system from observations}\label{sec:anest}

\subsection{Constraining the binary orbit using astrometric data}\label{sec:binorbastrom}

\begin{figure*}
    \centering
    \includegraphics[width=\textwidth]{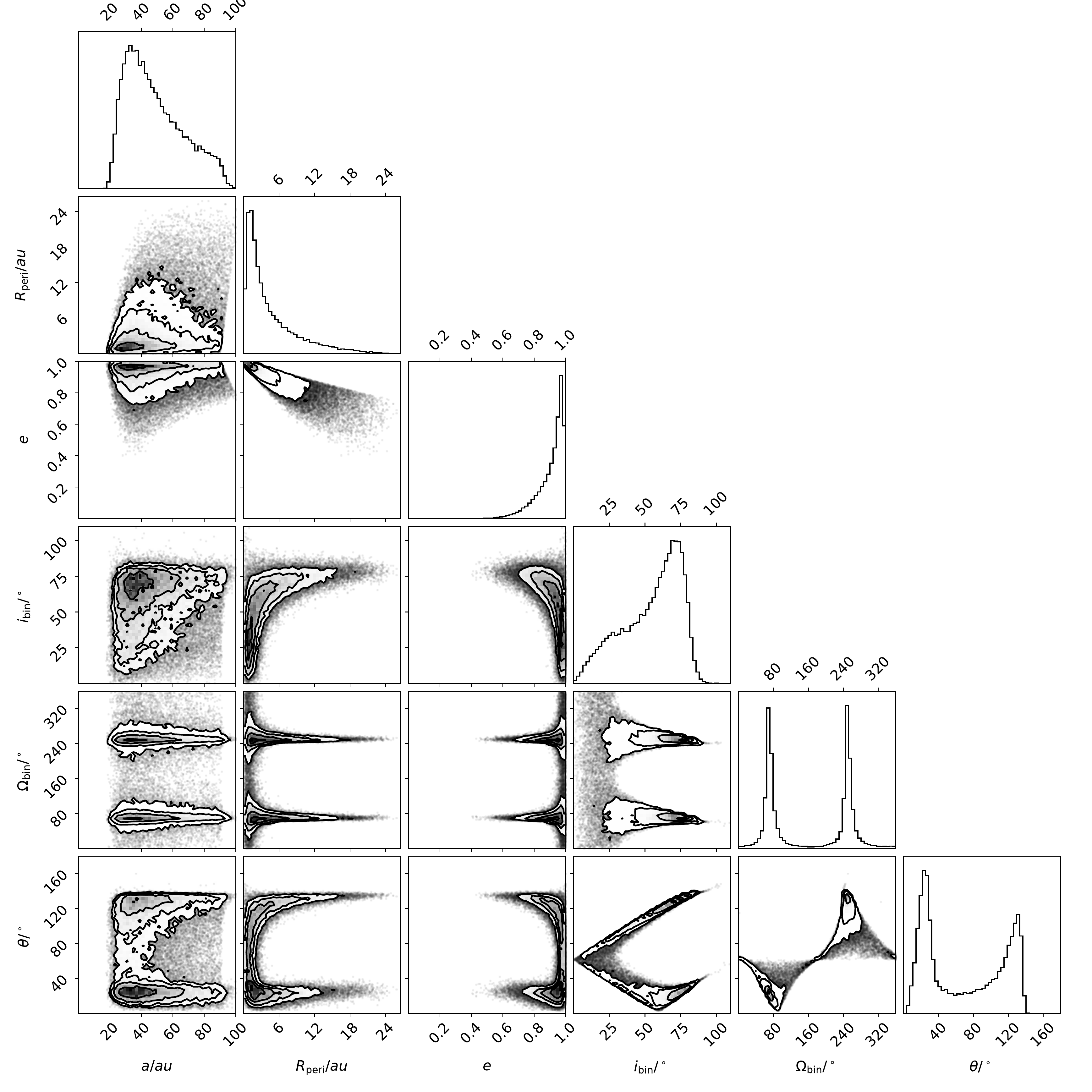}
    \caption{Results from \textsc{imorbel} \citep{pearce2015} fit of the astrometric data in Fig. \ref{fig:keck}. The histograms represent the posterior distributions of the orbital parameters fitting the astrometric data completed with $8\times10^4$ uniformly distributed couples $(z,v_z)$ for the variables $\{a,R_{\rm peri},e,i_{\rm bin},\Omega_{\rm bin},\theta\}$: i.e., semi-major axis, pericentre radius, eccentricity, inclination, longitude of the ascending node, mutual binary-disc inclination (as detailed in Eq. \ref{eq:thetaincl}, assuming $\Omega_{\rm disc}=93^\circ$ and disc inclination $i_{\rm disc}=62^\circ$). We assumed a total mass of the binary $M_{\rm \star, tot }=0.3\pm 0.1\,{\rm M_\odot}$. Each set of orbital parameters collected in the histogram represents an orbit consistent with the astrometric arc. Peaks in the distributions should not be interpreted as most reliable outcomes of the fitting procedure: these distributions are sensitive to the choice of the prior distribution of the couples $(z,v_z)$. }
    \label{fig:distrib}
\end{figure*}

We perform a fit of the astrometric data (shown in Fig. \ref{fig:keck}) using \textsc{imorbel} \citep{pearce2015}. For a given sequence of separations and PAs of the binary along an arc in the plane of the sky, \textsc{imorbel} interpolates the velocities and positions $(x,y,v_x,v_y)$ measured by the astrometric data and makes an assumption about z-coordinate and z-velocity, by sampling uniformly couples $(z,v_z)$, that cannot be captured due to the projection of the orbit along the line of sight of the observer; the code outputs the distributions of the orbital elements that best fit the arc.

We prescribe the binary mass to be $M_{\rm \star,tot}=0.3\pm 0.1\,{\rm M_{\odot}}$, as discussed in Sec. \ref{sec:stars}\footnote{The measure performed by \citet{andrews2008} and our independent mass estimate suggest $M^-_{\rm \star,tot}=0.15\,{\rm M_\odot}$ and $M^+_{\rm \star,tot}=0.45\,{\rm M_\odot}$ as ``extreme'' acceptable values; we adapted the mass errors to $M_{\rm err}=0.10\, {\rm M_{\odot}}$ in order to limit the mass sampling beyond $1.5\sigma=0.15\,{\rm M_{\odot}}$ when using  \textsc{imorbel}: Gaussian sampling of $M_{\rm mean}=0.3\, {\rm M_{\odot}}$ $M_{\rm err}=0.10\, {\rm M_{\odot}}$ will prevent values with $M<0.15\, {\rm M_{\odot}}$ and $M>0.45\,{\rm M_{\odot}}$ to be extensively used.}.
We set the fit to exclude those orbits that would end up with an apoastron distance $R_{\rm apo}>180$ au, as the maximum separation of the binary along its orbit cannot exceed the size of the cavity ($R_{\rm cav}\sim 185$ au).

We present in Fig. \ref{fig:distrib} the distributions of semi-major axis, pericentre separation, eccentricity, binary inclination, longitude of the ascending node and mutual binary-disc inclination: $(a,R_{\rm peri},e,i_{\rm bin},\Omega_{\rm bin},\theta)$, respectively.
The distribution of mutual binary-disc inclination angle $\theta$ (bottom line of Fig. \ref{fig:distrib}) is obtained using \citep{czekala2019}:
\begin{equation}
\cos(\theta)=\cos(i_{\rm disc})\cos(i_{\rm bin})+\sin(i_{\rm disc}) \sin(i_{\rm bin}) \cos(\Omega_{\rm disc} - \Omega_{\rm bin}),\label{eq:thetaincl}
\end{equation}
assuming $\Omega_{\rm disc}= 93^\circ$ and $i_{\rm disc}= 62^\circ$ (see Sec. \ref{sec:IRASdiscs}) and couples of $\{i_{\rm bin},\Omega_{\rm bin}\}$  from each fitted orbit.

It is important to note that these distributions should not be read as a statistical representation of the best fitting parameters, but as an indication to constrain the boundaries of the parameter space describing the orbit. Indeed, each $(z,v_z)$ couple produces a set of orbital parameters that represents the best fit for such a couple; the distributions simply quantify the number of $(z,v_z)$ couples that are consistent with a certain value of a given orbital element and highlight the correlations between the parameters. Different prior distributions of $(z,v_z)$ can substantially change the posterior distribution of the orbital elements \citep{pearce2015}.

With reference to Fig. \ref{fig:distrib}, the resulting distribution of the mutual inclination favours values of $\theta_1\sim 25-30^\circ$ for a binary ascending node $\Omega_{\rm bin}\sim 75^\circ$ and  $\theta_2\sim120\textrm{--}125^\circ$ for a binary ascending node $\Omega_{\rm bin}\sim 255^\circ$, for $i_{\rm bin}\approx 70$\degr.

More generally, the inclination of the orbit cannot exceed $i_{\rm bin}\gtrsim 75^\circ$ (note that a $i_{\rm bin}\approx 90^\circ$ is readily excluded by the fact that the orbital motion is not along the radial direction). We also note that inclinations $i_{\rm bin}\lesssim 50^\circ$ imply $R_{\rm peri}\approx 4$ au, which is in tension with the size of the circumstellar discs, as we will discuss in the next section.

Possible values of the binary eccentricity are  $e>0.5\textrm{--}0.6$ with a peak in the distribution at $e\approx 0.95$.
Even though for low-mass/large-period binaries the orbital eccentricity distribution is a flat curve (see Tab 13 in \citealp{moe2017}), making in fact eccentricities as high as $e=0.95$ physically meaningful, we anticipate that in the following discussion we will further constrain that the binary eccentricity should not exceed $e>0.7$. Lowest eccentricities appear to be available only for inclination values $i_{\rm bin}\sim 65^\circ\textrm{--}75^\circ$ (see correlation $e-i$ in Fig. \ref{fig:distrib}), further strengthening the reliability of our estimate of $\theta$.

Finally, concerning the binary semi-major axis, values of $a\sim 20\textrm{--}95$ au seem possible. Further considerations, taking into account the disc structure, are required to better constrain the value of $a$, as we will discuss in Sec. \ref{sec:binorbbindisc}.

\subsection{Constraining the binary orbit using binary-disc interaction theory}\label{sec:binorbbindisc}

It is possible to obtain more constraints by considering how the binary shapes the disc.

We first compare cavity and circumstellar disc sizes with theoretical predictions for given inclinations and eccentricities. The binary is expected to truncate the disc through two possible mechanisms: resonant \citep{artymowicz1994,miranda2015} and non-resonant \citep{papaloizou1977,rudak1981,pichardo2005,pichardo2008}.
Both approaches to disc truncation have been developed for gaseous discs, even though the results from \citealp{rudak1981}, \citealp{pichardo2005} and \citealp{pichardo2008} about non-resonant truncation were based on calculations with ballistic particles. \citet{dipierro2017} found that planetary companions are more efficient at opening gaps in the dust rather than in the gas; also, disc external truncation by a companion has been observed to occur at smaller radii in the dust than in the gas (\citealt{manara2019}, Rota et al., submitted). However, we are not aware of analytical works for large mass-ratio companions studying tidal internal/external truncation of the dust component. In the following discussion we will rely on the assumption that dust and gas are sufficiently coupled to have a similar dynamics. Our predictions will be tested numerically in Sec. \ref{sec:SPH}.

The general trend for both theoretical frameworks is that the ``truncation efficiency'' grows for growing binary eccentricities and/or decreasing mutual inclinations between the disc and the binary orbital plane $\theta$. This implies that in general, for a fixed value of $\theta$ and binary semi-major axis, larger values of binary eccentricity will produce larger cavities and smaller circumstellar discs. Vice versa, for a fixed binary eccentricity and semi-major axis of the binary, the co-planar case ($\theta=0^\circ$) will always produce the largest cavity and smallest circumstellar discs.
We are currently not aware of truncation models accounting also for the circumbinary disc eccentricity; some numerical results have been discussed in \citet{ragusa2020} for circular binaries, but there are no elements that could allow us to generalise those results to eccentric binaries. In the light of this, in our analysis we prescribe that the truncation radius is equal to the cavity semi-major axis ($R_{\rm cav}=a_{\rm cav}$) and apply theoretical results developed for circular discs.

Regarding the possible values of mutual inclination (see Sec. \ref{sec:binorbastrom}), on the one hand, the low mutual binary-disc inclination of the case with $\theta=25\textrm{--}30^\circ$ produces results similar to the co-planar case. For this case, resonant theory predicts a sharp transition from a cavity truncation radius of $R_{\rm cav}\sim 2a_{\rm bin}$ (resonance 1:3) to $R_{\rm cav}\sim 3a_{\rm bin}$ for binary orbital eccentricities $e_{\rm bin}\gtrsim 0.5$ -- i.e., truncation occurs at the 1:5 resonance (\citealp{miranda2015}, see their Fig. 9). The exact threshold value of the binary eccentricity for this transition depends on the disc viscosity, and inclination $\theta$. In any case, we expect a semi-major axis of $\sim 60$ au to be able to create a cavity of $\sim 185$ au for any value of binary eccentricity $e\gtrsim 0.5$ and $\theta\lesssim 25^\circ$.

On the other hand, the high inclination and retrograde nature of the orbit with $\theta\sim 125^\circ$, implies that the binary truncation cannot occur at radii larger than $R_{\rm cav}\sim 2a_{\rm bin}$ for any choice of binary eccentricity \citep{miranda2015}. These considerations imply that for the high inclination case a binary semi-major axis of $a\sim 100$ au is necessary to truncate the circumbinary disc at the observed radius. This high inclination configuration can be excluded:
large mutual inclinations between the binary and disc do not favour the formation of an azimuthal over-dense feature, which appears only for relatively low inclination discs ($\theta\leq 30^\circ$, \citealp{poblete2019}); furthermore, a retrograde binary would appear as an inversion of the velocity field in the cavity region of moment 1 map (see Fig. \ref{fig:MmapsObs}) due to the material in the circumstellar discs orbiting in the opposite direction with respect to that in the circumbinary disc, which is not the case. These considerations make a binary-disc inclination of $\theta_2\sim 125^\circ$ unlikely to properly reproduce the morphology of this system.

In order to be more quantitative regarding the semi-major axis, we analyse the low inclination case using non-resonant truncation theory \citep{pichardo2005,pichardo2008}, which predicts a smoother function of $R_{\rm cav}(a,e)$ than the typical step-like shape for resonant truncation. Despite some differences, both theories produce comparable predictions.
In this framework, the cavity size $R_{\rm cav}$ for a given binary semi-major axis ($a$), mass ratio ($q$) and binary eccentricity ($e$) is given by \citep{pichardo2008}
\begin{equation}
    R_{\rm cav}(a,e,q)=1.93\cdot\left(1+1.01\cdot e^{0.32}\right)\left[\frac{q}{(1+q)^2}\right]^{0.043} a,\label{eq:cavsize}
\end{equation}
where $q=M_2/M_1$.

As previously mentioned in Sec. \ref{sec:IRASdiscs}, the inferred cavity size from the observations is $\sim 185$ au at the source distance of $D\sim 130$ pc.

\begin{figure}
    \centering
    \includegraphics[trim={0.3cm 0 1.3cm 0.5cm},clip,width=\columnwidth]{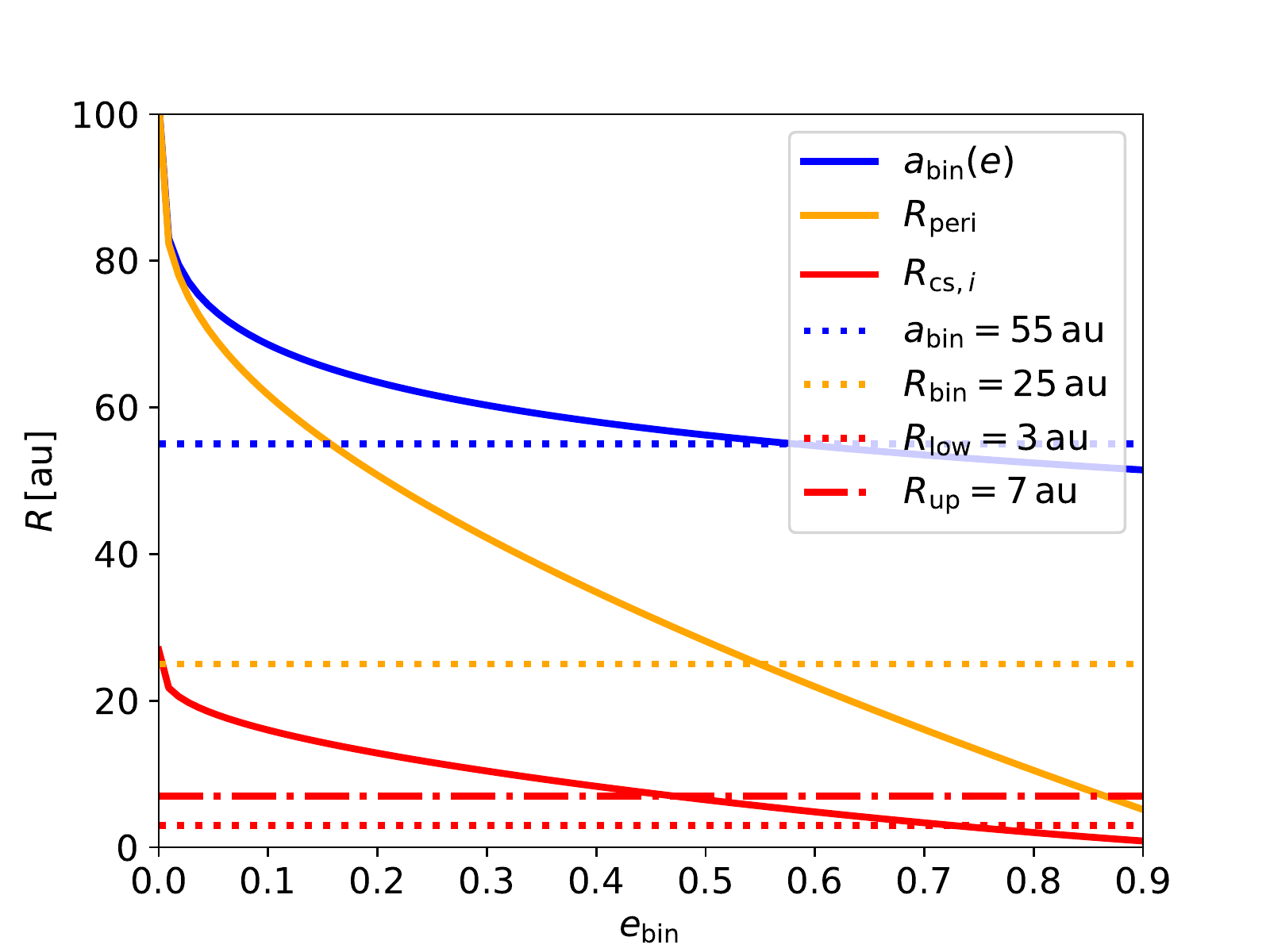}
    \caption{The blue line shows the degeneracy between the binary semi-major axis $a_{\rm bin}$ and its eccentricity to obtain a cavity $R_{\rm cav}=185$ au, using Eq. (\ref{eq:cavsize}). Based on the blue curve, orange and red curves show the dependence of pericentre radius ($R_{\rm peri}=a_{\rm bin}[1-e_{\rm bin}]$) and external truncation radius of circumstellar discs ($R_{{\rm cs},i}$, from Eq. \ref{eq:truncrad}) as a function of eccentricity, respectively. Blue, orange and red dotted lines are meant as reference values for binary semi-major axis ($a_{\rm bin}=55$ au, to be compared with blue curve) projected binary separation ($R_{\rm bin}\approx 25$ au, to be compared with orange curve) and half of the beam minor-axis in Band 4 ($R_{\rm low}\approx 3$ au and $R_{\rm up}\approx 7$ au, to be compared with red curve), respectively. }\label{fig:avse}
\end{figure}

We show in Fig. \ref{fig:avse} the couples of binary orbital parameters $\{a,e\}$ with which an equal mass binary will carve a cavity of $\approx 185$ au in the circumbinary disc (blue curve), using Eq. (\ref{eq:cavsize}). From this figure, it is easy to notice that binary eccentricity values $e>0.5$ predicted from the fit of the astrometric data, intrinsically exclude values of semi-major axes $a> 55$ au.

Despite not constraining directly the binary orbit, we note that the relatively small projected separation $R_{\rm proj}=0\farcs18$ of the binary ($\sim 25$ au at the source distance) does not favour the largest values of semi-major axis, which are associated to binary eccentricity $e\lesssim0.5$: indeed, the pericentre radius $R_{\rm peri}=a(1-e)$ (orange curve) would not be sufficiently small to allow such a projected separation at any point of the binary orbit.

Analogously, we can use the size of the circumstellar discs to put an upper-limit on the binary eccentricity. As previously mentioned in Sec. \ref{sec:binorbbindisc}, dust circumstellar discs are marginally resolved in Band 4 (zoom-on panel of Fig. \ref{fig:contB4B7obs}), but are most likely to be inclined with respect to the plane of the sky consistently with the beam shape, so that $R_{{\rm cs},i}\gtrsim 3\, {\rm au}$ constitutes a lower limit for their radius. Using the non-resonant truncation radius equation by \citet{pichardo2005}\footnote{Results of resonant truncation for co-planar circumstellar discs in equal mass binaries (see Fig. 5 in \citealp{miranda2015}) appear to be in good agreement with the non-resonant prediction by \citet{pichardo2005}.}
\begin{equation}
   R_{{\rm cs},i}= 0.733\cdot(1-e)^{1.2} \left(\frac{q}{1+q}\right)^{0.07}R_{i,{\rm Roche}},\label{eq:truncrad}
\end{equation}
where $R_{i,{\rm Roche}}$ is the \citet{eggleton1983} estimate of the Roche radius of the $i$-th star of the binary
\begin{equation}
    R_{i,{\rm Roche}}=\frac{0.49\cdot q_i^{2/3}}{0.6\cdot q_i^{2/3}+\ln\left(1+q^{1/3}\right)}a_{\rm bin},
\end{equation}
and $q_1=M_1/M_2$ and $q_2\equiv q=M_2/M_1$, we obtain $R_{{\rm cs},i}\sim 3\, {\rm au}$ for $e\sim 0.7$. This result places an upper limit on the binary eccentricity that must satisfy $e\lesssim 0.7$ for the circumstellar discs to have $R_{{\rm cs},i}\gtrsim 3\, {\rm au}$.

\begin{figure}
    \centering
    \includegraphics[trim={0cm 4.5cm 0cm 5.5cm},clip,width=\columnwidth]{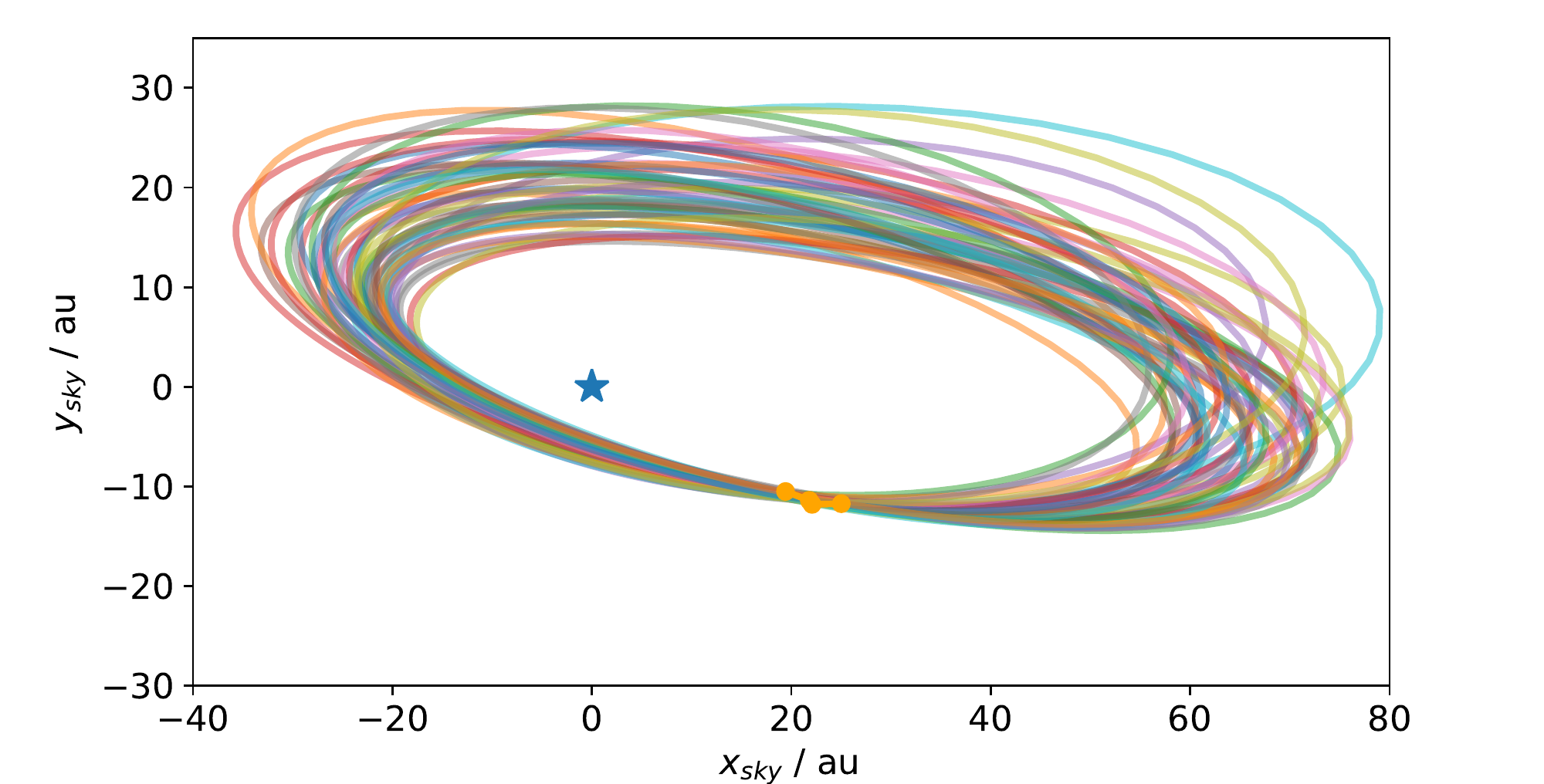}
    \caption{Selection of 50 orbits fitting the astrometric measures of the binary in Fig. \ref{fig:keck}. Such orbits satisfy the additional constraints discussed in Sec. \ref{sec:anest}, namely: binary eccentricity $0.5<e<0.7$, semi-major axis $50<a<65$ au and inclination $50^\circ<i<75^\circ$. The measured astrometric arc is marked in orange. We refer to Fig. \ref{fig:keck} for a zoom-on the arc area. }
    \label{fig:orbsel}
\end{figure}

In summary, both the astrometry, presented in Sec. \ref{sec:binorbastrom}, and the disc size study in this section support a binary with mid-to-high eccentricity $e\approx 0.5\textrm{--}0.7$, moderate binary-disc misalignment $\theta\approx 30^\circ$ and a binary semi-major axis $a\approx 50\textrm{--}60$ au. A selection of orbits from the \textsc{imorbel} fit satisfying these constraints is shown in Fig. \ref{fig:orbsel}.

\subsection{Constraining dust and gas masses from observations}\label{sec:dustmasses}
Under the assumption that the dust is optically thin, we can derive the total dust mass contained in the disc from the total flux $F_\nu=259\pm 26$ mJy in Band 7 (336 GHz) using \citep{beckwith1990}
\begin{equation}
    M_{\rm dust}\approx \frac{D^2F_\nu}{\kappa_\nu B_\nu(20\,K)}\approx 1.5\times 10^{-4} M_\odot,\label{eq:dmass}
\end{equation}
where $D\sim130$ pc is the distance from the source,  $\kappa_\nu=1.97\cdot(\nu/336\,{\rm GHz})^{0.4}\, {\rm cm^2\,g^{-1}}$ is the dust opacity, $B_\nu(20\, K)$ is the Planck's black body emission assuming an average dust temperature of $\langle T\rangle\sim 20$ K \citep{andrews2008,long2018}. The value $M_{\rm dust}\approx 1.5\times 10^{-4}\, {\rm M_\odot}$ is consistent with previous estimates ($M_{\rm dust}\sim 2\times 10^{-4}\, {\rm M_\odot}$ \citealp{glauser2008,andrews2008}).
Using a typical gas-to-dust ratio $M_{\rm g}/M_{\rm d}\approx 100$, our dust mass estimate leads to a total gas mass of $\sim 20\, {\rm M_J}$. We note that a direct measure of the gas-to-dust ratio $M_{\rm gas}/M_{\rm dust}=220^{+150}_{-170}$ has been provided by \citet{glauser2008}, thanks to a separate measurement of the gas mass (fitting X-ray emission) and dust mass along the line-of-sight to the star. However, the dust mass estimate was made by assuming a complete disc orbiting a single star. For this reason, and in view of the new data available indicating a different geometry, especially on the line-of-sight to the center, caution should be used when applying the gas-to-dust ratio derived by \citet{glauser2008}.

Using the same approach provided in Eq. (\ref{eq:dmass}), we present also a mass estimate of the circumstellar discs. We assume an average temperature of circumstellar discs to be $\langle T\rangle\sim 50$ K \citep{vanderplas2016}.
We get that each circumstellar disc has a dust mass $M_{{\rm B7},i}\approx M_{{\rm B4},i}\sim 1.3\times 10^{-6}\,{\rm M_\odot}$, based on the total flux from the individual discs in Band 7 and Band 4 ($F_{{\rm B7},i}$ and $F_{{\rm B4},i}$ in Tab. \ref{tab:flux}), respectively.
However, we note that both the temperature and the assumption that the material is optically thin in the circumstellar discs might not be satisfied, making in fact any estimate of their mass based on Eq. (\ref{eq:dmass}) only a lower limit.

\section{Numerical Simulations}\label{sec:SPH}

In this section we further constrain through numerical simulations the subset of orbital configurations of the binary that better reproduce the dynamics and electromagnetic appearance of the system, further refining the cohesive model we have developed in Sec. \ref{sec:anest}. With this goal in mind, based on the previous theoretical estimates, we perform a set of 3D SPH gas + dust numerical simulations using the code \textsc{Phantom} \citep{price2018a}.

Then, starting from the dust density distribution in the hydrodynamic simulations, we carry out a set of Monte Carlo radiative transfer simulations using the code \textsc{Mcfost} \citep{pinte2006,pinte2009} in order to directly compare the expected electromagnetic output of the simulations with the observations.

We note that the numerical simulations in this work are not meant to provide a fine tuning of the disc and binary parameters, nor to study the long term evolution of the system.
As a consequence, we will not perform a systematic study of the disc structure exploring the parameter space. We will also limit the study of our systems to the first 50 binary orbits of evolution, when both circumstellar discs are still present in the simulation -- after that time, mostly for numerical reasons, their accretion rate exceeds the rate at which they are fed from the edge of the cavity, which causes them to disappear at long timescales (see further discussion in Sec. \ref{sec:results}).

\subsection{Hydrodynamic simulations}\label{sec:numsim}

We use the 1-fluid algorithm to treat dust dynamics \citep{LaibePrice2014a,PriceLaibe2015,ballabio2018} implemented in \textsc{Phantom} \citep{price2018a}, which is valid under the assumption of strong coupling between dust and gas (Stokes number ${\rm St}<1$). In this formalism, one single set of particles stores the information regarding both the gas component and the dust one. Each particle represents a fluid element of the gas + dust mixture with a fixed total mass; the amount of gas and dust carried by each particle is described by the dust-fraction $\epsilon$, which is allowed to evolve as well as all the other standard hydrodynamic quantities.

The binary is modeled using two sink particles of mass $M_1=M_2=0.15\, {\rm M_\odot}$, for a binary total mass $M_{\rm \star,tot}=0.3\, {\rm M_\odot}$, consistently with what is discussed in Sec. \ref{sec:stars}. Gas + dust particles are removed from the simulation and considered accreted when they cross the sink radius $R_{\rm sink}=1$ au following the prescription in \citet{bate1995}. The sink particles are allowed to freely move under their mutual gravitational attraction and the back-reaction force the disc exerts onto the binary.

We use SPH artificial viscosity to model the physical mechanisms  transporting the angular momentum through the disc by prescribing an artificial viscosity $\alpha_{\rm AV}=0.4$, which results in a \citet{shakura1973} viscosity parameter $\alpha_{\rm ss}\simeq 0.005$ \citep{lodato2010}; we also use $\beta_{\rm AV}=2$ to prevent particle interpenetration \citep{price2018a}. We use a locally isothermal equation of state: we prescribe the sound speed to follow a radial power law profile $c_{\rm s}=c_{s0}(R/R_{\rm ref})^{-0.25}$, where $R_{\rm ref}=100$ au and $c_{\rm s0}$ chosen in order to have a disc aspect ratio $H/R=0.2$ at $R_{\rm ref}$, consistent with the parameters derived in Sec. \ref{sec:IRASdiscs}. Such a choice is consistent with a temperature profile $T\propto R^{-1/2}$, typical of passive protostellar discs.

Our disc consists of a set of $N_{\rm part}=3\times 10^6$ particles initially distributed between $R_{\rm in}=50$\,au and $R_{\rm out}=700$\,au in a way that the initial surface density profile is a tapered radial power law $\Sigma=\Sigma_0(R/R_{\rm ref})^{-p}\exp[-(R/R_{c})^{2-p}]$, with $p=1$, $R_{\rm ref}=100$\,au, $R_{\rm c}=600$\,au.
$\Sigma_0$ is set so that the total mass of the disc is $M_{\rm disc}=0.02\,M_{\odot}$, consistent with what derived in Sec. \ref{sec:dustmasses}.
With an initial inner radius $R_{\rm in}<50$\,au (or a full disc), the material in the cavity would be evacuated anyway but at a greater computational cost.

The dust component is set prescribing an initial gas-to-dust ratio of $100$ uniformly across the entire disc domain. We decide to use dust particles of $140\, {\mu m}$ in size, representative of the dust mixture because they are the ones with the maximum cross-section for emission in Band 7, at $890 {\rm \mu m}$.

We use a binary with semi-major axis $a=55$ au, as discussed in Sec. \ref{sec:binorbastrom} and \ref{sec:binorbbindisc}.
With such a choice of parameters the binary is initially embedded in the disc, so that the cavity and circumstellar discs can adjust their size after the simulation starts.
Considerations about the binary eccentricity in Sec. \ref{sec:binorbbindisc} suggest that its  value should be $0.5<e<0.7$. We choose $e=0.5$ for our simulations. Larger binary eccentricities result in a rapid depletion of the circumstellar discs. We postpone a more detailed discussion about this effect to Sec. \ref{sec:discussion}.
We simulate three binary-disc inclinations of $\theta=\{0^\circ,30^\circ,60^\circ\}$, in order to test the agreement of the numerical simulations with the prediction of low-moderate $\theta$ we made in Sec. \ref{sec:binorbbindisc}.
At the end of the setup procedure, positions and velocities of particles and sinks are adjusted in order to have the centre of mass of the whole system and its velocity set to $\bm x_{\rm CM}=(0,0,0)$ and $\bm v_{\rm CM}=(0,0,0)$.

We let the system evolve for $t=50$ binary orbits, which corresponds to a physical time of $\sim 4\times 10^4$ yr. The reason for this choice will become more clear in the following sections; in short: numerical effects prevent the circumstellar discs to be long-lived.
We also note that the numerical setups presented in this section are extremely expensive computational-wise.
Indeed, they require a large dynamic range --- from a few au (circumstellar discs) up to hundreds of au (circumbinary disc) --- and for this reason they require a high spatial resolution. They are 3D, as we need to account for inclination. We must consider dust dynamics. Evolving each simulation for $50$ orbits requires $\approx60$k cpu hours (i.e., more than 2 months of uninterrupted computation on a 36 CPU machine).

\subsection{Radiative transfer and synthetic images}\label{sec:radtransfer}

In order to directly compare the results in our hydrodynamic simulations with the observations, we create Band 4/Band 7 continuum images and CO moment 0 and moment 1 maps of our simulations using the Monte Carlo radiative transfer code \textsc{mcfost} \citep{pinte2006,pinte2009}. \textsc{mcfost} uses as an input the distribution of gas and dust grains to create a Voronoi mesh, each cell corresponds to an SPH particle. The disc receives passive heating from the radiation of the two central stars, which are here modelled as two black bodies with temperature $T=3080$ K (consistently with their stellar mass $M_{\star,i}=0.15\,{\rm M_\odot}$ assuming an age of $1\textrm{--}3$ Myr, \citealp{siess2000}) located at the position of the sinks. We note that \citet{andrews2008} were not able to measure the luminosity of the stars due to large uncertainties in the extinction. This, coupled with the imprecise kinematic mass estimate of the binary, implies an uncertainty on the radius, temperature and luminosity of the stars, which might vary significantly depending on their age.

We assume a DIANA standard dust composition \citep{woitke16a,min16a} -- i.e. 60\% silicates, 15\% amorphous carbon (optical constants from \citealp{dorschner1995} and \citealp{zubko1996}, respectively) and 25\% porosity. Opacities were then computed assuming a minimum dust grain size $s_{\rm min}=0.03 \,{\rm \mu m}$ and maximum $s_{\rm max}=3\,{\rm mm}$ and a grain distribution $dn(s)/ds\propto s^{-7/2}$.
The code assumes that grains smaller than 1 ${\rm \mu m}$ follow the gas, the spatial distribution of dust grains between 1 ${\rm \mu m}$ and 140 ${\rm \mu m}$ is obtained by interpolation. Dust grains with size $140\,{\rm \mu m}<s<3$ mm have the spatial distribution of the dust component in the numerical simulations.

We discretise the radiation field using $N_{\gamma}=10^7$ photons in order to compute the temperature structure of the disc assuming local thermal equilibrium (LTE), images are then computed via ray-tracing again with $N_{\gamma}=10^7$ photons. We compute Band 4 (145 GHz, 2.06 mm), Band 7 (336GHz, 0.890 mm) continuum images, and CO maps using the assumption $T_{\rm gas}=T_{\rm dust}$.
Line emission refers to the CO molecular transition $J=3-2$, using a ratio $N_{\rm CO}/N_{\rm H}= 10^{-4}$, which is a commonly assumed value (e.g., \citealp{miotello2016}). The CO maps are obtained spanning velocities between $v_{\rm min}=-6\,{\rm km\,s^{-1}}$ and $v_{\rm max}=6\,{\rm km\,s^{-1}}$ using a channel width $\Delta v=0.03\,{\rm km\,s^{-1}}$ that were then post-processed to simulate the $\Delta v_{\rm obs}= 0.25\,{\rm km\,s^{-1}}$ of observations.
Since the gas is optically thick, the line flux is not sensitive to increase or reduction of the gas mass.

The theoretical dust mass estimate from Eq. (\ref{eq:dmass}) of $M_{\rm dust}=1.5\times 10^{-4}\,{\rm M_\odot}$ matches with the observed fluxes.

The fluxes were obtained accounting for the circumbinary disc inclination of $i=62^\circ$ with respect to the line of sight and smoothing the infinite resolution image with Gaussian beams consistent with those synthesized in the observations in Band 4, Band 7 and CO maps (see Sec.~\ref{sec:obs}).

We performed some tests including the CO freeze-out at temperatures $T<20$ K, however this produced a steep cut in the CO maps which is not visible in the observations (no emission at radii larger than $R\gtrsim 200$ au). The outer disc is possibly receiving heating from an external source or photo-desorption is effective. For this reason we decided not to include the effect in our models.

\subsection{Results}\label{sec:results}

\begin{figure}
	\includegraphics[trim={0.2cm 0cm 9cm 0cm},clip,width=0.5\textwidth]{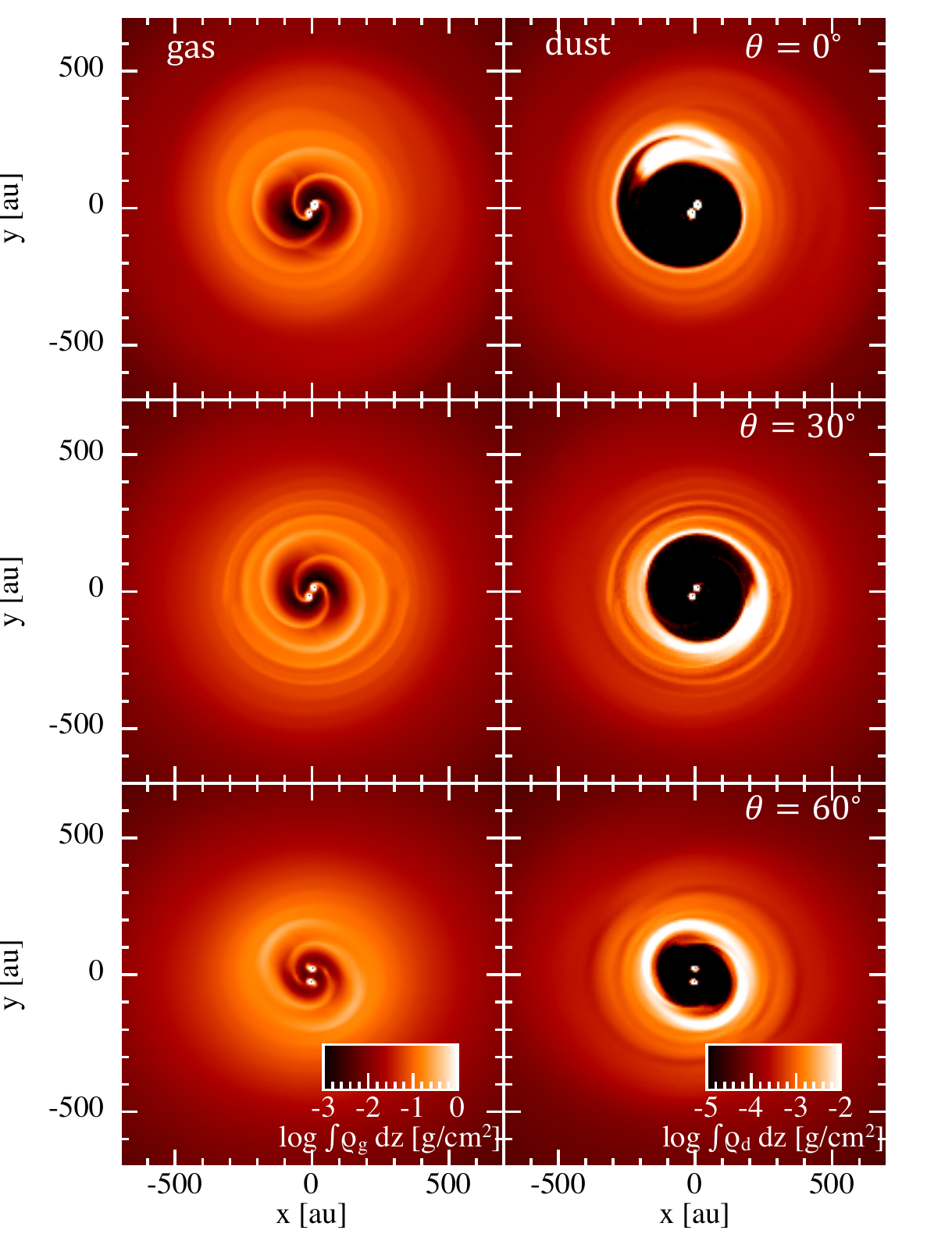}
    \caption{Gas and dust density snapshot from the simulations  after $\sim 40$ binary orbits ($\approx$30\,kyr). Each row shows a different mutual binary-disc inclinations $\theta=\{0^\circ,30^\circ,60^\circ\}$ (top, central and bottom row, respectively). Left panels show gas density, right panels show dust. }
    \label{fig:snapshots}
\end{figure}

\begin{figure}
    \centering
    \includegraphics[trim={0cm 0.2cm 1cm 1.5cm},clip,width=\columnwidth]{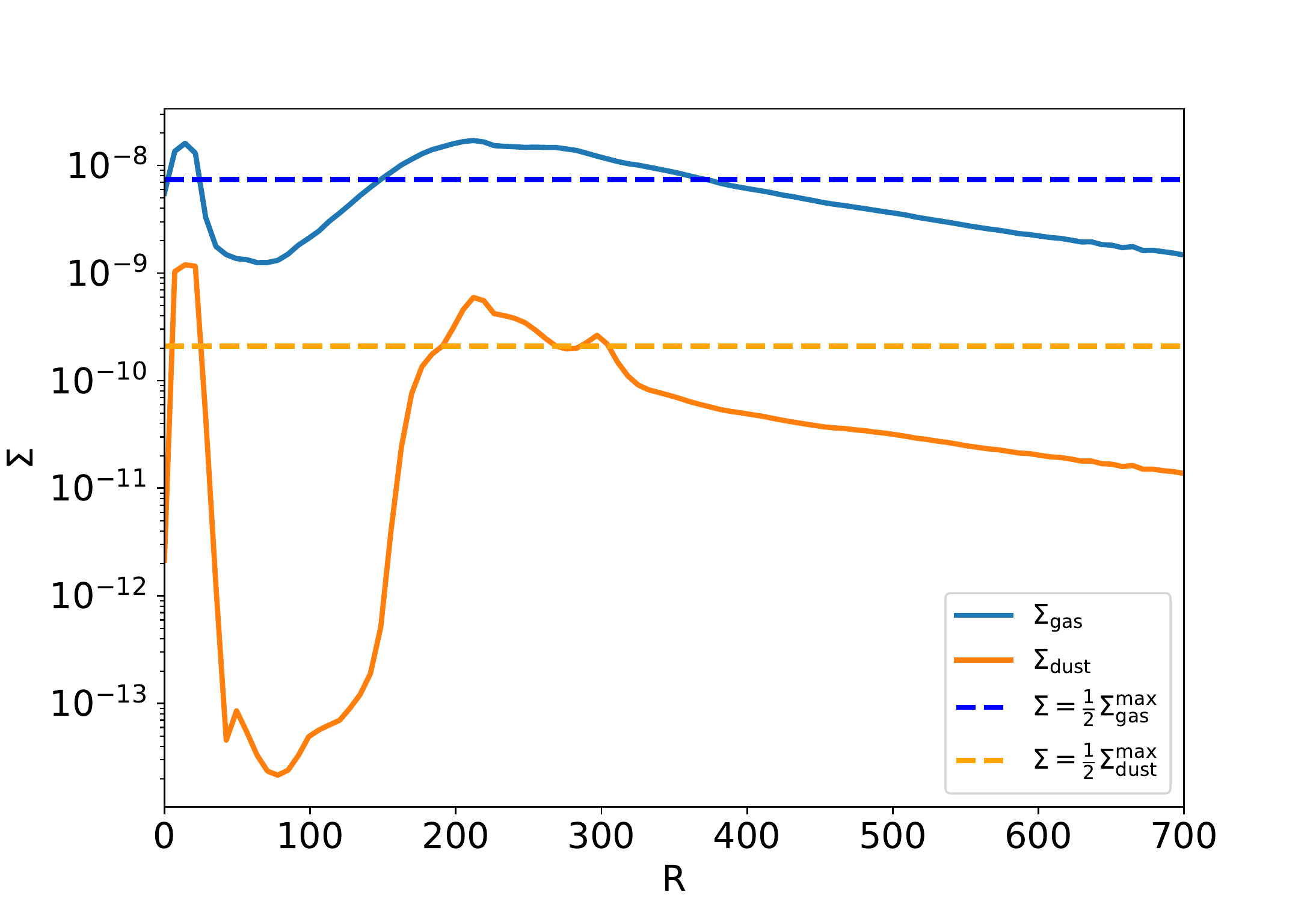}
    \caption{Gas (blue) and dust (orange) radial density profiles obtained by azimuthal average for the $\theta=0^\circ$ simulation. Dashed lines represent half of the peak density value.}
    \label{fig:densprof}
\end{figure}

As soon as the simulations start, the binary repels the material from its co-orbital region. After $\sim 40$ binary orbits ($\approx 3\times10^4$ yr)
the binary has carved a cavity in the gas, whose density profile reaches its peak at $R_{\rm peak}\sim 150$ au --- snapshots in  Fig. \ref{fig:snapshots}, see also the radial profile obtained from azimuthal averaging of the case $\theta=0^\circ$ in Fig. \ref{fig:densprof}. A large amount of gas keeps flowing through the cavity in the form of gas tidal streams feeding the binary, while the dust grains form a steeper dust cavity of $R_{\rm cav}\sim 190$ au. Two circumstellar discs of radius $R_{{\rm cs},i}\sim 10$ au form around the individual elements of the binary. These discs appear to be aligned with the orbital plane of the binary.

For moderate inclinations, $\theta=0^\circ\textrm{--}30^\circ$, the binary also produces a prominent azimuthal over-density that follows the Keplerian motion of the flow at the cavity edge. No azimuthal over-density is observed in the $\theta=60^\circ$ case. Such an over-density is most evident in the dust density field, where it reaches a contrast ratio $\sim 10$; in contrast, in the gas density field it is barely visible.

In our $\theta>0$ simulations the disc shows a tendency to align with the binary orbital plane: at the time of the snapshot in Fig. \ref{fig:snapshots}, in the $\theta=30^\circ$ case, the binary and the disc have a mutual inclination $\theta_{\rm snap}=25^\circ$; in the $\theta=60^\circ$ case, the binary and the disc have a mutual inclination $\theta_{\rm snap}=55^\circ$.

The circumstellar discs are characterised by an unphysically high accretion rate. Their limited spatial resolution\footnote{Note that in SPH resolution depends on the fluid density. High density implies a large number of particles in a limited space region and thus a high resolution,} produces a spurious growth of the artificial viscosity and triggers a positive feedback loop (i.e. accretion rate increases as they become sparser) that causes them to accrete on the sinks at a rate progressively faster than that at which they are fed from the edge of the cavity. This implies that after $\sim 50$ binary orbits from the beginning of the simulation the circumstellar discs completely disappear.
In reality, the circumstellar discs are expected to reach a quasi-steady state when the accretion rate on the individual stars $\dot M_{\star,i}$ equals the rate at which they are fed from the edge of the cavity $\dot M_{\rm cav}$.
Since the accretion rate on the individual stars depends on the surface density $\Sigma$ as $\dot M_{\star}\propto \nu\Sigma$, one can always find a value $\Sigma=\Sigma_{\rm steady}$ such that $\dot M_{\star}\propto \nu\Sigma_{\rm steady}\approx \dot M_{\rm cav}$.

In the light of these considerations, we caution that the amount of material in the two circumstellar discs does not represent a reliable estimate of the real amount of material available surrounding the two stars.

\begin{figure*}
    \includegraphics[trim={1.7cm 2cm 2.2cm 3.7cm},clip,width=0.7\textwidth]{8902060obs.png}
	\includegraphics[trim={1.7cm 2cm 2.2cm 3.7cm},clip,width=0.7\textwidth]{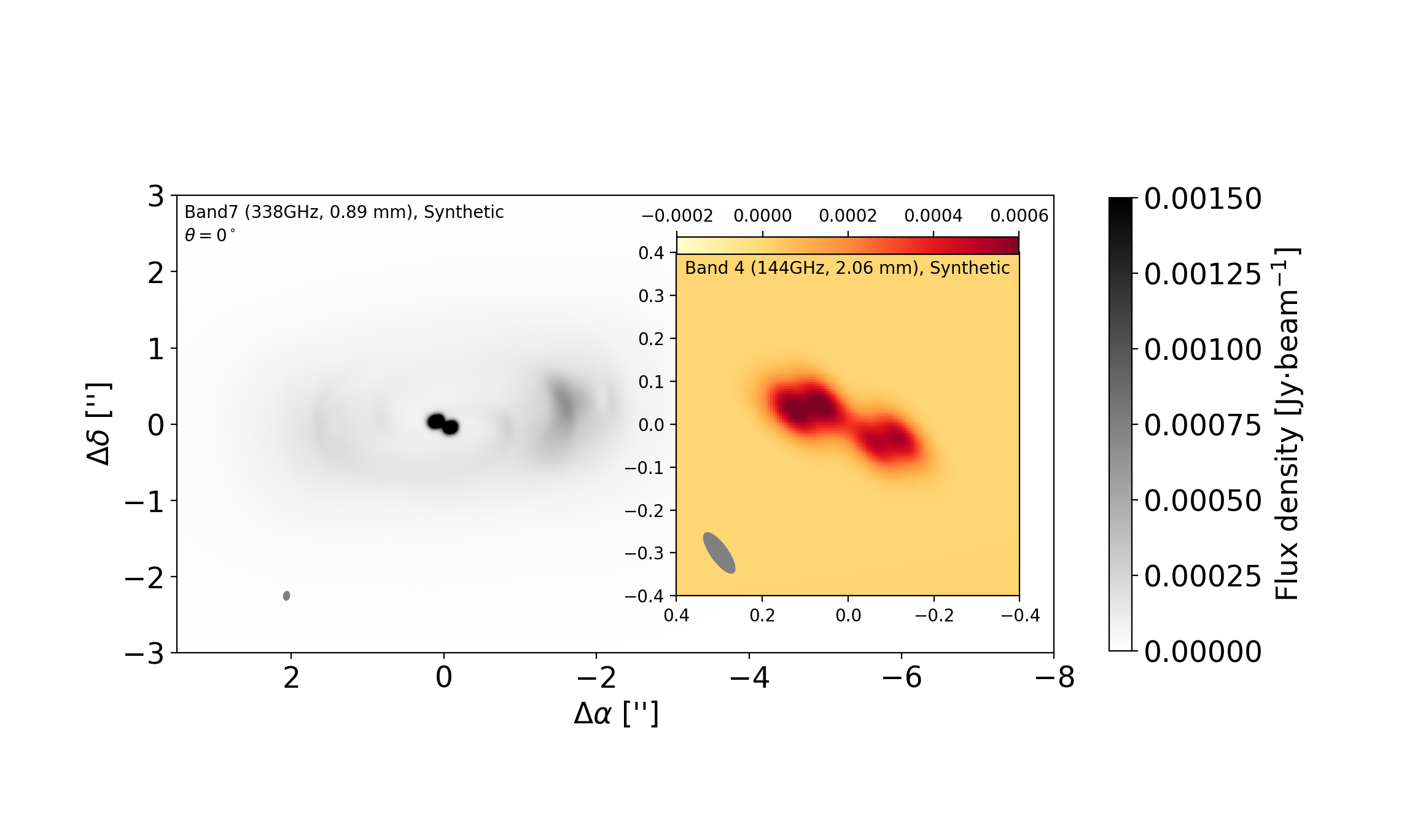}
	\includegraphics[trim={1.7cm 2cm 2.2cm 3.7cm},clip,width=0.7\textwidth]{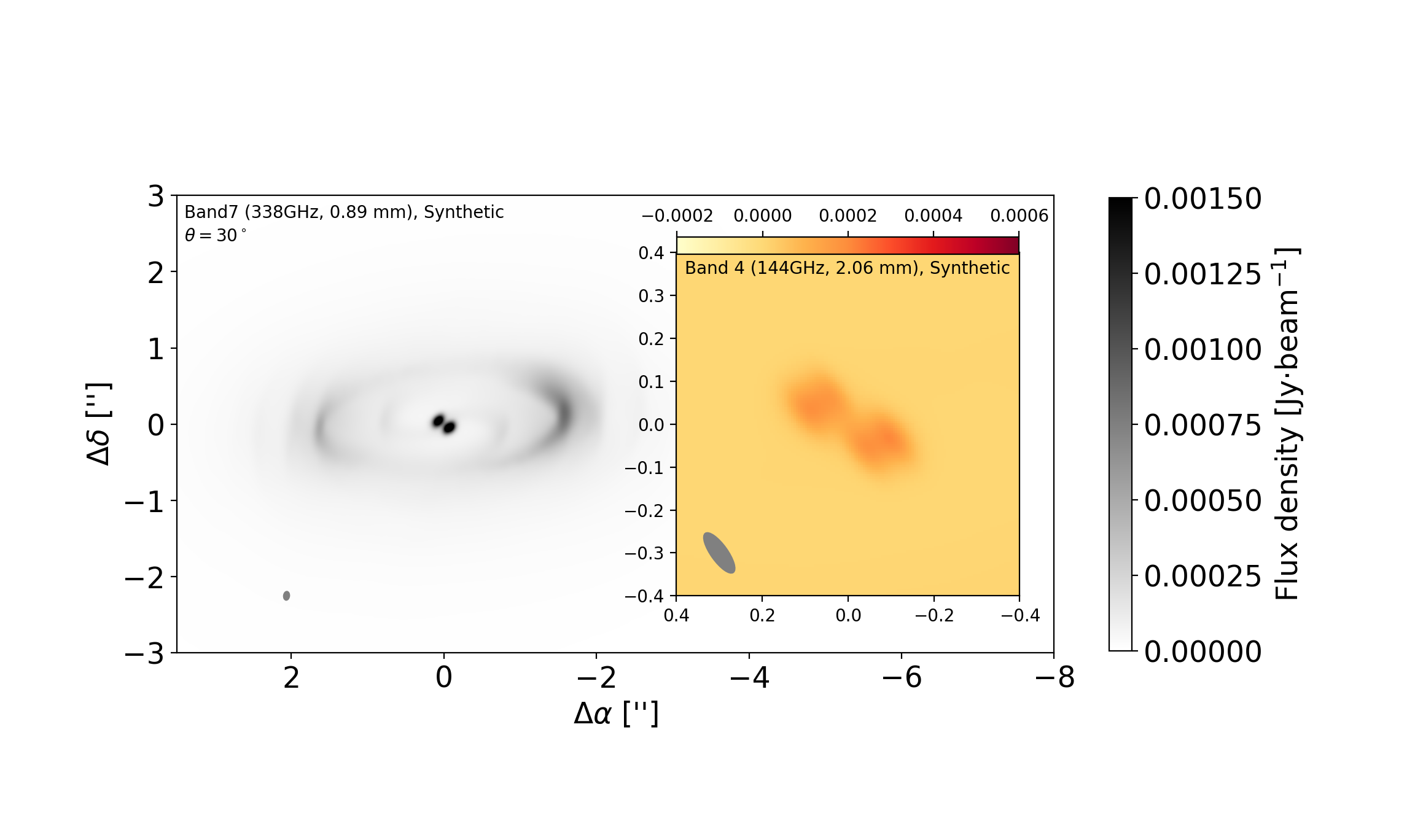}
	\includegraphics[trim={1.7cm 2cm 2.2cm 3.7cm},clip,width=0.7\textwidth]{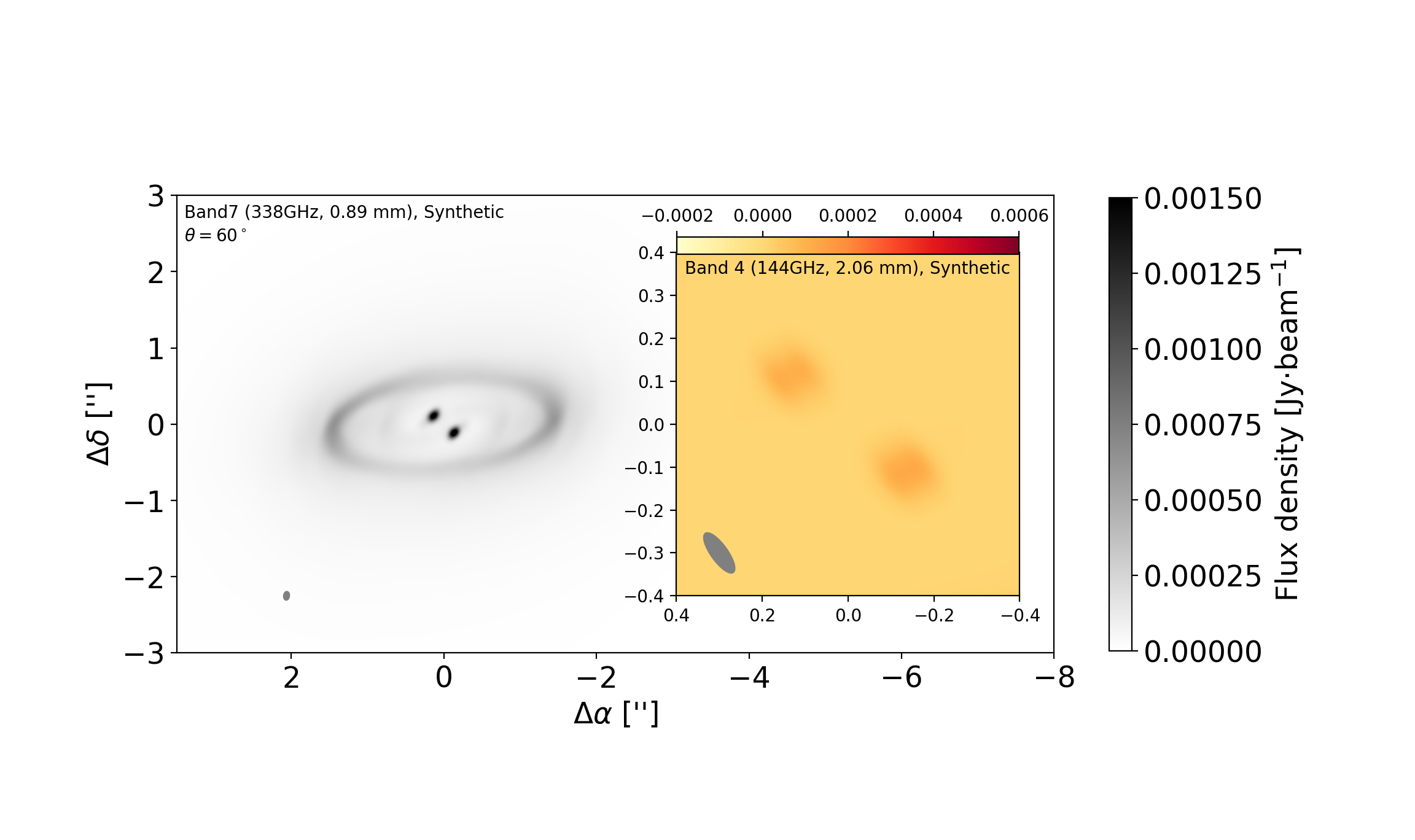}
    \caption{Comparison between ALMA Band 7 (compact + extended configuration) and Band 4 continuum observations (top row, originally presented in Fig. \ref{fig:contB4B7obs}) and our synthetic models for binary-disc mutual inclinations $\theta=\{0^\circ,30^\circ,60^\circ\}$ (row 2, 3 and 4, respectively), obtained applying Monte Carlo radiative transfer post-processing of the simulation snapshots shown in Fig. \ref{fig:snapshots}. Grey-shaded plots show Band 7 extended array configuration, while the zoom on the circumstellar discs (orange-shaded panel) shows Band 4.  Note that the apparent ``source doubling'' in synthetic Band 4 images are artefacts caused by the sink particles (having a minimum in brightness at their center) and the elongated shape of the beam.
    }
    \label{fig:continuum2}
\end{figure*}

Fig. \ref{fig:continuum2} shows the snapshots presented in Fig. \ref{fig:snapshots} post-processed with radiative transfer, as described in Sec. \ref{sec:radtransfer}; the images combine Band 4 ($2060\,{\rm \mu m}$, zoom-in plot, orange-shaded colour-scale) and Band 7 ($890\,{\rm \mu m}$, main plot, grey shaded colour scale), to be compared with the actual ALMA observations in the top row --- previously shown in Fig. \ref{fig:contB4B7obs}.

\begin{figure*}
    \includegraphics[trim={4.cm 0.3cm 2.7cm 1.8cm},clip,width=0.49\textwidth]{COM0obs.png}
	\includegraphics[trim={4.cm 0.3cm 2.7cm 1.8cm},clip,width=0.49\textwidth]{COM1obs.png}\\
    \includegraphics[trim={4.cm 0.3cm 2.7cm 1.8cm},clip,width=0.49\textwidth]{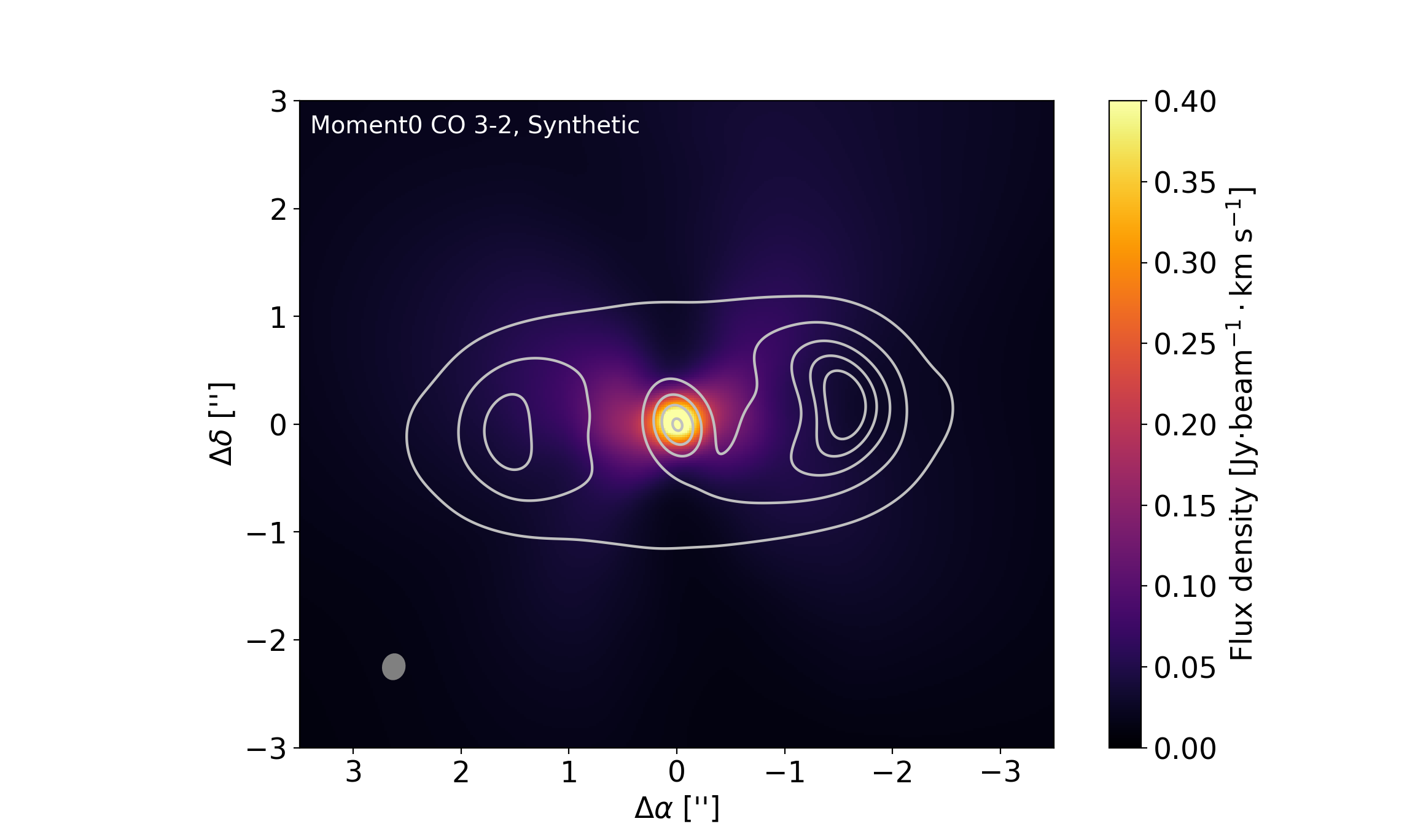}
	\includegraphics[trim={4.cm 0.3cm 2.7cm 1.8cm},clip,width=0.49\textwidth]{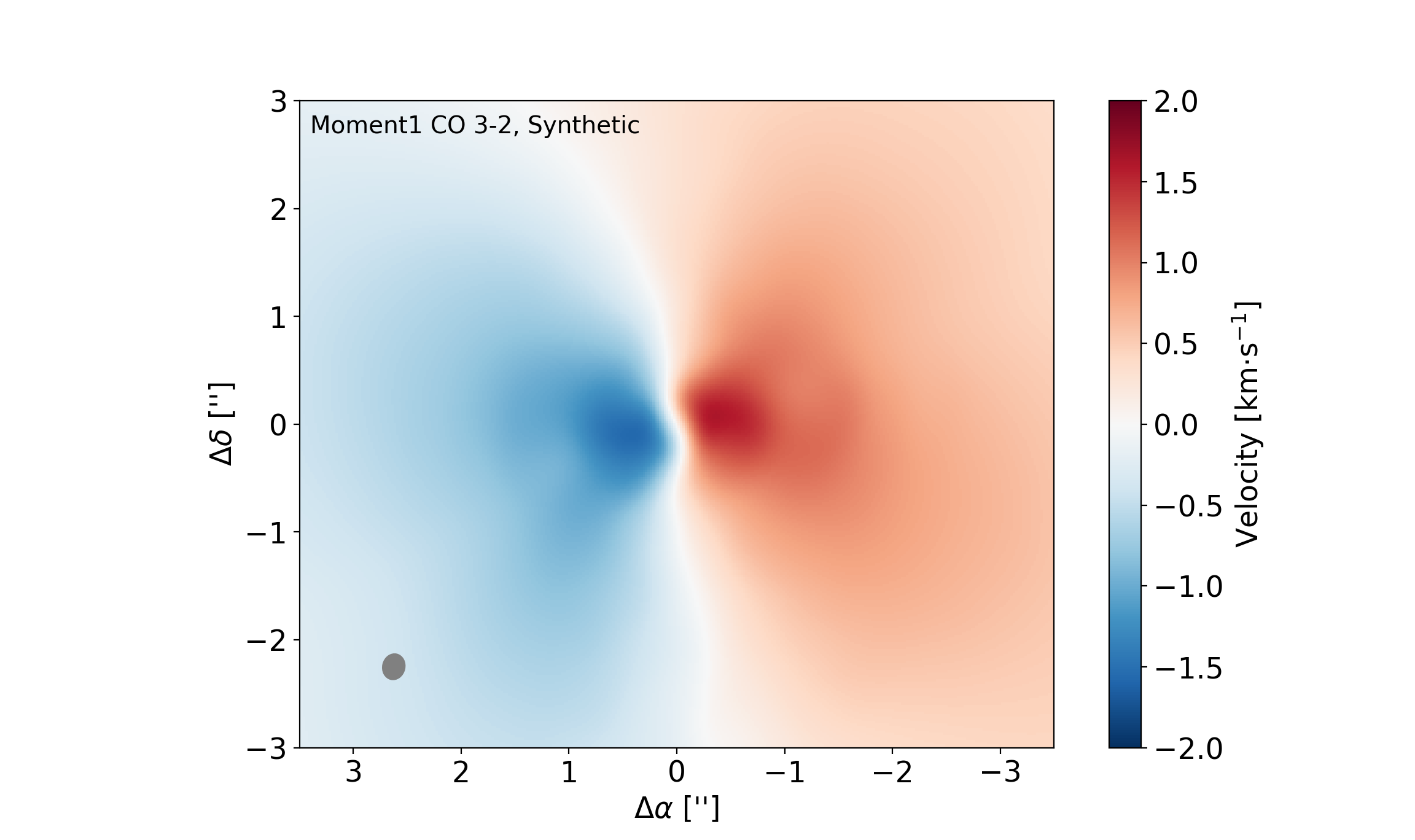}\\
    \caption{Comparison between observations of $J=3\textrm{--}2$ CO moment 0 (left panel) and moment 1 (right panel) maps (top row, originally presented in Fig. \ref{fig:MmapsObs}). Synthetic model for binary-disc mutual inclinations $\theta=30^\circ$ (bottom row), obtained from Monte Carlo radiative transfer post-processing of simulation snapshot in Fig. \ref{fig:snapshots}. Contours superimposed to the moment 0 maps show emission levels of flux density $I_\nu=\{1.5,3.,5.,7.,9.\}$ mJy from the compact array configuration data only, in order to exclude short spatial wavelengths.}
    \label{fig:COmaps}
\end{figure*}

We show in Fig. \ref{fig:COmaps} synthetic moment 0 and moment 1 maps for the case $\theta=30^\circ$, again from the snapshots shown in Fig. \ref{fig:snapshots}.
We note that our models of CO line emission show very similar features for all binary-disc mutual inclinations $\theta$, making in fact impossible to rely on them to distinguish the most suitable model. However, we consider it instructive to show that qualitative agreement between simulations and observations concerning flux of the moment 0 map and morphology of the velocity field of moment 1 (S-shape of $v_{\rm proj}=0$ region) is achieved.
Synthetic moment 0 and moment 1 maps were obtained using the \texttt{Line.plot\_map} function of \textsc{pymcfost}\footnote{\url{https://github.com/cpinte/pymcfost}}. Contours in the synthetic moment 0 maps were obtained from the same Band 7 ($890\,{\rm \mu m}$) radiative transfer models presented in Fig. \ref{fig:continuum2}, this time smoothed with a beam  consistent with the ALMA compact configuration, for a direct comparison with the contours in the observed moment 0 map -- previously shown in Fig. \ref{fig:MmapsObs}.

\section{Discussion}\label{sec:discussion}

In the following discussion we will analyse in detail how the results in our simulations compare with the theoretical estimates (provided in Sec. \ref{sec:anest}) and with the observations of IRAS~04158+2805.

\subsection{Dust mass}

Our radiative transfer continuum models (see Fig. \ref{fig:continuum2}) match the observed fluxes in Band 7 using a total dust mass $M_{\rm dust}= 1.5\times10^{-4}\, {\rm M_\odot}$, consistent with our estimate from Sec. \ref{sec:binorbbindisc} using Eq. (\ref{eq:dmass}). \citet{glauser2008} previously estimated the dust mass $M_{\rm dust}=1\textrm{--}1.75\times 10^{-4}\, {\rm M_\odot}$, i.e. in perfect agreement with our models. \citet{sheehan2017} estimated the dust mass $M_{\rm dust}\approx1.5\times 10^{-3}\,{\rm M_\odot}$, which is one order of magnitude larger than the dust mass we used for our models. Using a typical gas-to-dust ratio $M_{\rm gas}/M_{\rm dust}=100$, such an estimate would lead to a total gas mass comparable to the mass of the binary, which should result in the disc being gravitationally unstable.
If that was the case, such a disc would show typical spiral features and possibly other kinematic effects, which we do not see \citep{hall2020,veronesi2021,paneque2021}.

Fluxes from the circumstellar discs in ALMA Band 4 radiative transfer model (zoom-on orange-shaded panel in Fig. \ref{fig:continuum2}) appear to be in good agreement for the case $\theta=0^\circ$ and $\sim 2$ times less luminous than the observed ones for $\theta>0^\circ$.
We note that the dust mass contained in the circumstellar discs at the moment of the snapshot for the cases $\theta>0^\circ$ is $M_{{\rm dust/sim},i}\approx 5\times 10^{-7}\, {\rm M_\odot}$, which is a factor $\sim 2$ lower than the estimate we provided in Sec. \ref{sec:dustmasses}. This discrepancy is qualitatively consistent with the lower brightness from the circumstellar discs.
In any case, we recall that the mass of the circumstellar discs are probably underestimated as discussed in Sec. \ref{sec:results}, and for this reason it should not be used when comparing the models with observations.

We conclude this section with a word of caution when comparing the models with the observations. Even though an overall good internal consistency between the theoretical estimates (Eq. \ref{eq:dmass}) and synthetic fluxes is obtained within the paper, it is important to note that estimates of the dust mass from the total luminosity are strongly affected by the underlying model (Class II \citealp{glauser2008}; Class I \citealp{sheehan2017}) and assumptions on the dust opacity $\kappa_\nu$: changing the porosity, dust grain size and composition might significantly alter the estimated mass.

The size of the discs and whether their emission is optically thick (in particular for circumstellar discs) have an impact on the total luminosity. Finally, variations in the stellar properties, which are to some degree uncertain, affect the disc flux.

\subsection{Gas CO line emission}\label{sec:COdiscuss}

The moment 0 map of the CO line emission for the transition $J=3-2$ from our radiative transfer model (left panel of bottom row in Fig. \ref{fig:COmaps}) qualitatively matches the observed one (Fig. \ref{fig:MmapsObs}). Despite the gas depletion in the cavity area surrounding the binary, enough material is still available to provide optically thick line emission from the centre of the system.
This is not obvious. Indeed, we note that preliminary tests with smaller disc aspect ratios $H/R\sim 0.1$ -- we recall our simulations use $H/R\sim 0.2$ -- showed lack of gas in the centre of the system. As a consequence, synthetic CO moment 0 maps presented a cavity in the gas as well as in the dust for relatively thinner discs. This result appears to confirm that the disc thickness is correctly represented in our modeling of the system.

However, our moment 0 map does not match the flux scale: in particular, our radiative transfer model is $\sim 1.5$ times fainter than the observed one.
Such a discrepancy might suggest that the innermost material in our simulations is slightly colder than in the observations, implying that our models require slightly larger stars in order to precisely match the observed fluxes. This suggests that the stars in IRAS~04158+2805 are possibly younger than 1 Myr or slightly more massive than we assumed.
Another possible origin of this discrepancy could be our assumption that $T_{\rm gas}=T_{\rm dust}$ when performing the radiative transfer. Above the $\tau=1$ optical depth surface, $T_{\rm gas}$ will start decoupling from $T_{\rm dust}$, becoming significantly hotter in the upper layers of the disc atmosphere. That temperature difference could compensate the missing flux in our models.

Although we did not manage to match the details of the gas emission, our synthetic moment 0 map captures well the strong emission from the cavity area.
Additional constraints on the stellar emission properties are required in order to improve the modeling of the gas emission, however overall we believe our models provide a decent match with the observations.

We defer the discussion about the gas kinematics and moment 1 maps to Sec. \ref{sec:evo}.

\subsection{Circumbinary and circumstellar discs morphology}

\begin{figure}
    \centering
    \includegraphics[trim={3cm 0.5cm 2.9cm 1cm},clip,width=\columnwidth]{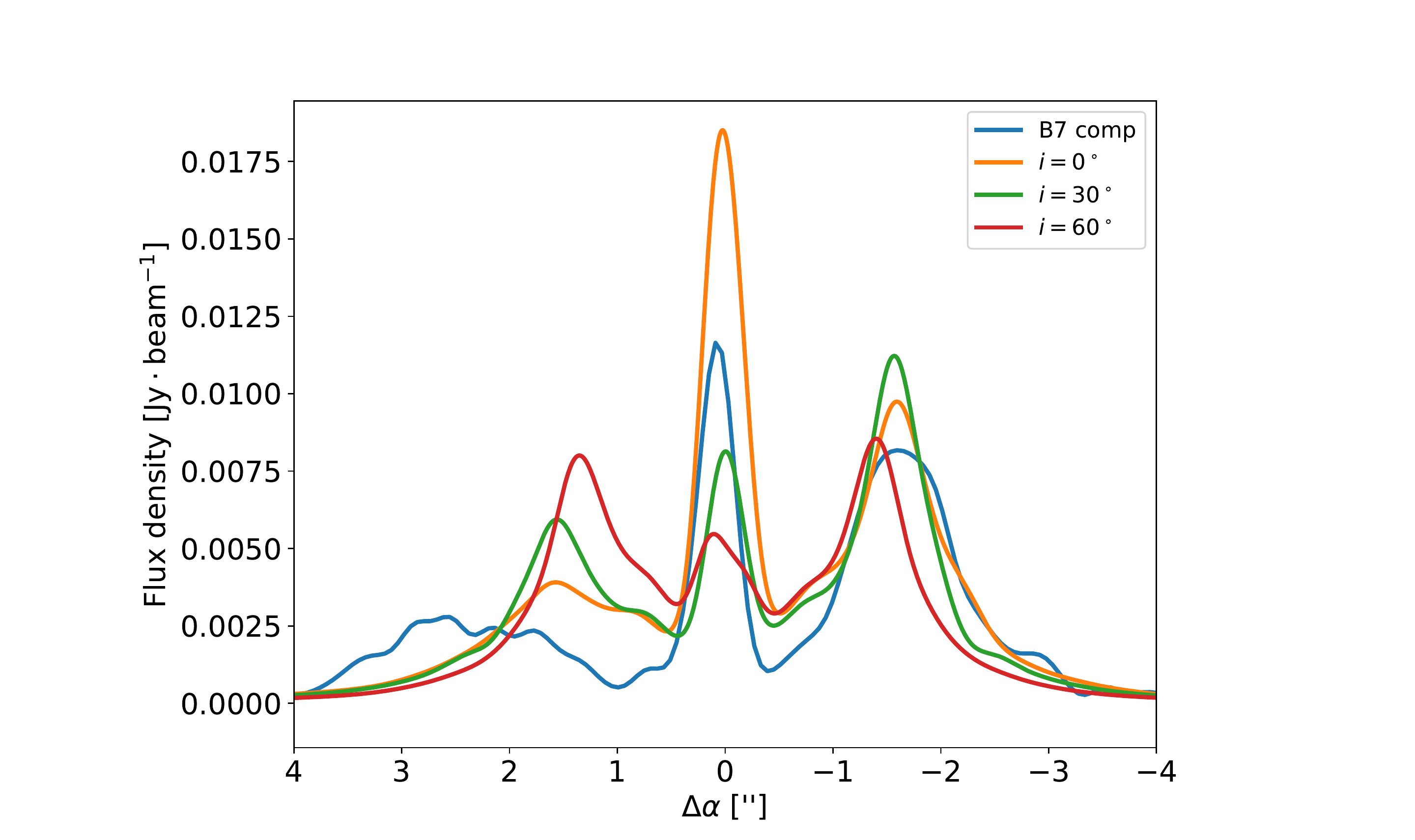}
    \caption{Cuts along the major axis of the cavity from Band 7 compact configuration images (see Fig. \ref{fig:cutsObs}). Different colours represent observations (blue line) and synthetic observations with $\theta=\{0^\circ,30^\circ,60^\circ\}$ (orange, green and red lines, respectively).
    }
    \label{fig:cuts}
\end{figure}

With reference to Fig. \ref{fig:snapshots}, all our simulations show a cavity size $R_{\rm cav}\sim 3a_{\rm bin}$. Fig. \ref{fig:cuts} shows a comparison of cuts along the cavity major axis from both observations and synthetic images (Fig. \ref{fig:contB4B7obs} and \ref{fig:continuum2}).
All simulations qualitatively match the observed cavity semi-major axis $a_{\rm cav}\sim 185$ au. Even though, $a_{\rm cav}$ slightly decreases for growing mutual binary-disc inclination $\theta$, as expected (see Sec. \ref{sec:binorbbindisc}). In contrast, simulations appear to fail to reproduce the observed cavity eccentricity. In particular, directly measuring the apocentres and pericentres of the dust cavities from (Fig. \ref{fig:continuum2} and profiles in Fig. \ref{fig:cuts}) the synthetic images of our models suggests the dust cavity is in fact consistent with being circular ($e_{\rm cav}\lesssim 0.05$), to be compared with $e_{\rm cav}\sim 0.2$ of the observed one. However, we note that dust cavities in the snapshots of our models (Fig. \ref{fig:snapshots}) show a certain degree of lopsidedness. We quantified the cavity eccentricity comparing apocentre and pericentre in the snapshots of $\theta=0^\circ\textrm{--}30^\circ$ simulations, we obtained $a_{\rm cav}\sim 190$ au and $e_{\rm cav}\sim 0.2$. The difference with the observations appears then to be mainly due to the fact that it was not possible for us to find a snapshot where the phase of the binary, the azimuthal position of the over-density and longitude of the cavity pericentre were right in place at the same time. The gas snapshots instead show a symmetric gas cavity -- which is consistent with the absence of asymmetric features in the moment 1 map and P--V diagram (Fig. \ref{fig:MmapsObs} and \ref{fig:pvdiag}). We note that we are not aware of any theoretical attempt to model the mutual disc eccentricity evolution in the gas and dust component, which could be an interesting starting point for additional physical considerations in future analyses.

Only the two moderate binary-disc inclination cases ($\theta=\{0^\circ,30^\circ\}$) reproduce the asymmetric emission observed at the cavity edge. The co-planar case, in particular, appears to perfectly reproduce the morphology of the western azimuthal over-dense feature with a contrast $\delta_\phi\sim 4$. We note that the binary in the $\theta=60^\circ$ was slightly farther from pericentre when the snapshot was taken than the other cases; the phase of the binary is not determining the presence/absence of the over-density: the over-density is not present at any orbital phase of the binary.

In our simulations circumstellar discs with radius $R_{{\rm cs},i}\sim 10$ au form immediately after the beginning of the simulation, consistently with the value expected for the eccentricity of our binary ($e=0.5$) from truncation theory (see Sec. \ref{sec:binorbbindisc}).
A close inspection of the circumstellar discs in the radiative transfer models (Fig. \ref{fig:continuum2}, in particular in the $\theta=0^\circ$ case) reveals a double lobed nature. We identify it as an artifact, a spurious feature at the centre of the sink, at radii smaller than sink particle size $R_{\rm sink}=1$ au\footnote{Note that the depleted area is larger than the sink diameter (i.e., $\approx 5$ au). The sink introduces an effective boundary condition on the density $\Sigma(R_{\rm sink})=0$, which causes the circumstellar disc density profile to decrease smoothly starting from a distance larger $R=R_{\rm sink}$.}, and because the convolution beam is very elongated.

We note that we performed some preliminary tests with binaries with $e=0.7$ with the goal to possibly observe smaller circumstellar discs. Such a large eccentricity configuration leads to a rapid numerical depletion of the circumstellar area and for this reason we decided to discard them, as they were not suitable for a direct comparison with the observations.
However, these simulations only showed a moderate reduction of the circumstellar disc size below $R_{{\rm cs},i}=10$ au ($R_{{\rm cs},i}=9$ au), which depends also on the phase of the binary.

As previously discussed in Sec. \ref{sec:binorbbindisc} (see also orange-shaded panel of Fig. \ref{fig:contB4B7obs}), circumstellar discs are marginally resolved in Band 4 observations (beam minor axis $\sim 3$ au), implying two main possible explanations of their apparent size.
First, their size is in fact $R_{{\rm cs},i}\sim 3$ au. This can be possibly explained by the fact that gas truncation radii are different from dust ones, with the latter being on average a factor 2--3 smaller than in the gas (\citealp{manara2019}, Rota et al. submitted). This makes in fact a gaseous disc of $R_{{\rm cs},i}^{\rm gas}\sim 10$ au, qualitatively consistent with a dust circumstellar disc with $R_{{\rm cs},i}\sim 3$ au. This result might indeed not be captured in our numerical simulations due to the limited resolution of the circumstellar discs.
Second, the observed circumstellar discs have in fact $R_{{\rm cs},i}\approx 10$\,au but they are inclined with respect to the plane of the sky and oriented accordingly with the highly elliptic beam. This effect is not captured in our numerical simulations but, if that was the case, their apparent size would be qualitatively consistent with the upper limit we discussed in Sec. \ref{sec:IRASdiscs}.

Misaligned circumstellar discs with respect to the circumbinary one could in principle cast shadows visible in scattered light images \citep{nealon2020}. However the high inclination of the circumbinary disc with respect to the line of sight and the large vertical extent of the disc ($H/R=0.2$) at large radii significantly limit our ability to access the inner regions of the disc at IR wavelengths, possibly preventing their detection. Furthermore, if the central binary is not co-planar with the circumbinary disc, then the circumstellar discs will also be out of plane and may / will not cast shadows.

\subsection{Kinematics and evolution of the system properties}\label{sec:evo}

The moment 1 maps in the right panel of Fig. \ref{fig:COmaps} show a good  agreement with the observed one. The scale of the velocity in our synthetic maps matches well the observed one, confirming our choice for the binary masses. The locus of points where $v_{\rm proj}=0$ is characterised by a peculiar ``S'' shape in both our simulated map and in the observed one. This effect could be due either to the ``warped'' nature of the cavity induced by the misaligned binary \citep{rosenfeld2014} or to strong radial motions due to lack of centrifugal balance. We note that a similar velocity pattern has been observed also in IRS48, where an inner binary has not been detected but it is strongly suggested by the disc structure \citep{calcino2019}.

Long term evolution of the cavity properties, i.e. size and eccentricity, has been documented in numerical works \citep{thun2017,hirsh2020,ragusa2020}. Results suggested that the gas cavity profile progressively steepens to match the theoretical prediction \citep{hirsh2020}. A large disc thickness (we recall in this case $H/R=0.2$) and viscosity $\alpha_{\rm ss}$ appear to prevent the disc eccentricity to reach large values in the gas \citep{ragusa2020}, as it would be instead expected for lower disc aspect ratios \citep{thun2017,miranda2017}. For this reason, we do not expect the cavity appearance in our numerical simulation to significantly change at longer timescales.

Recent theoretical \citep{aly2015,zanazzi2018,lubow2018}, numerical \citep{martin2018a,cuello2019a} and observational \citep{kennedy2019} developments have demonstrated that discs might tend towards polar alignment when they are surrounding highly eccentric binaries. It is possible to identify a critical inclination for polar alignment\footnote{Not to be confused with the $I_{\rm crit}$ parameter in \citet{zanazzi2018} (see their Eq. (8)), which is a necessary condition for polar alignment to occur, i.e. the minimum $\theta_{\rm crit}$ for a given binary eccentricity and $\Omega_{\rm e-d}=90^\circ$.}, $\theta_{\rm crit}(e,\Omega_{e-d})$, that depends on the binary eccentricity and on the angle between the longitude of the disc ascending node and binary longitude of pericentre $\Omega_{e-d}$ \citep{farago2010,cuello2019a}.
In our $\theta>0$ simulations the disc shows the tendency to align with the binary orbital plane towards a co-planar configuration. This is consistent with the theoretical expectations. The threshold for polar alignment is $\theta_{\rm crit}> 60^\circ$, when $e<0.7$ and $\Omega_{e-d}<20^\circ$, which appears to be the case for this system. As a consequence, we expect the disc in both our $\theta=30^\circ$ and $\theta=60^\circ$ simulations to become co-planar with the binary across a timescale of $\sim 1000$ binary orbits \citep{martin2018a}, i.e., a physical time of $\sim 10^6$ yr. In the light of the considerations in Sec. \ref{sec:obsdiscussion} regarding the spectral class of the system, we expect this system not to be older than $\sim 1\textrm{--}3$ Myr. Considerations concerning the luminosity of the central in Sec. \ref{sec:COdiscuss} and alignment of the circumbinary disc with the binary orbital plane appear to favour the younger side of this age range.

It is now an established numerical result that discs surrounding almost equal mass ratio binaries are characterised by a prominent azimuthal overdensity at the edge of the cavity \citep{shi2012,dorazio2013,farris2014,dorazio2016,ragusa2016,ragusa2017,miranda2017,ragusa2020}.
The survival of the azimuthal over-dense structure at the edge of the cavity has been documented after long timescales ($t\gtrsim 2000\,t_{\rm orb}$, \citealp{miranda2017,ragusa2020}). Such a density structure can be barely visible in the gas density field, nevertheless gas over-densities, even with contrast ratios as low as $\delta_\rho\sim 1.5$, that are co-moving with the flow can effectively trap dust grains creating high contrast dust over-densities \citep{calcino2019,poblete2019,vandermarel2021}.

\subsection{Origin of the azimuthal overdensity}

Azimuthal over-dense structures have been often observed in discs with cavities
(e.g. \citealp{andrews2011,isella2013,vandermarel2016,boheler2017,vandermarel2018,casassus2018,pinilla2018,vandermarel2019}).
Three main mechanisms for their formation have been suggested in the literature.
First, the formation of a vortex when the Rossby wave instability is triggered \citep[RWI,][]{lovelace99,ataiee2013,li2000,lovelace2014}. In that case, viscosity must be low (\citealp{shakura1973}, viscosity parameter $\alpha_{\rm ss}\sim 10^{-5}$).
Second, numerical simulations of discs surrounding high mass-ratio binaries often host a prominent azimuthal over-density at the edge of the cavity \citep{shi2012,dorazio2013,farris2014,dorazio2016,ragusa2016,ragusa2017,miranda2017,ragusa2020}. In this scenario, these structures appear to form regardless of the disc viscosity~--~such a feature is often present in numerical simulations of discs surrounding black hole binaries, where the viscosity parameter can be as high as $\alpha_{\rm ss}=0.1$. The formation of such a feature appears to be related to the tidal streams ``thrown'' by the binary against the cavity edge \citep{shi2012} and/or possibly to an unstable mechanism that is triggered when the cavity becomes eccentric \citep{ragusa2020}.
Third, the interaction of the disc with a (sub)stellar companion causes the disc eccentricity to grow. The clustering of the disc eccentric orbits produces an ``eccentric feature'', sometimes referred to as ``eccentric traffic jam'',
that appears as an azimuthal over-density at the cavity apocentre \citep{ataiee2013,teyssandier2016,thun2017}.

The first and second mechanism produce an azimuthal over-density that co-moves with the flow, i.e., it moves at the cavity edge with Keplerian frequency ($\sim 5$ binary orbital periods), and are expected to act as azimuthal dust traps \citep{vandermarel2021}. The third mechanism, because the over-density is always located at the apocentre of the eccentric cavity, produces an over-density that moves at the precession rate of the eccentric cavity, i.e., much slower than Keplerian ($\sim 100\textrm{--}1000$ binary orbital periods). Such an over-density does not selectively trap marginally coupled dust grains \citep{birnstiel13a}.

In the light of these considerations, we note that observing the over-density at the pericentre of the cavity -- as observed for IRAS~04158+2805 -- intrinsically excludes the ``eccentric traffic jam'' scenario, where the over-density is supposed to remain still at the apocentre of the cavity. Observing the over-density moving on a timescale $\ll 100$ binary orbits, or selective trapping of marginally coupled dust grains would confirm that the origin of such features is the first or second scenario we described. Measuring any type of vortical motion of the flow in the over-density region would put low upper limits on the disc viscosity ($\alpha_{\rm ss}\lesssim 10^{-5}$) and suggest that RWI is effective.

\subsection{Origin of the wiggling jet}

Young stellar sources are often characterised by the presence of supersonic collimated jets -- typical velocities range between $50\textrm{--}200\, {\rm km\,s^{-1}}$ (e.g., \citealp{podio2021} for a recent survey) --  launched from the close proximity of the central stars.

Fig. \ref{fig:hstwiggle} highlights the presence of a wiggling jet of material launched from the cavity area.
We first note that the counter jet, seen to the South of the system in the lower panel of Fig. \ref{fig:hstwiggle}, is also bent and appears to be the flip-mirror (rather than point-symmetric) counterpart to the main jet and, for this reason, is not consistent with a simple nozzle model.

With the aim to constrain the jet launching region, we proceed as follows:
we assume that the jet wiggling nature is due to some dynamical periodic oscillations with a characteristic frequency within the cavity region; we calculate typical oscillation frequencies; then, by multiplying the spatial extension of a wiggling segment times this frequency, we can calculate the launch velocity and compare it with typical wind velocities.

The distance travelled by the outflow in half wiggling cycle is $\sim 8''$, i.e. $\sim 10^3$ au. We first consider the nodal precession timescale of the circumbinary disc, which is $\sim 2\times 10^3$ binary orbits, i.e. $1.6$ Myr (Eq. 28 in \citealp{zanazzi2018}).
This makes the velocity of the outflow required to cover the spatial extent of $\sim 10^3$ au, $v_{\rm jet}\approx 3\,{\rm ms^{-1}}$, too slow for this type of jets. Using the same approach, but using the nodal precession rate of the circumstellar discs $t_{\rm prec, i}\approx 20\, t_{\rm orb}$ (Eq. 22 in \citealp{bate2000}) one gets $v_{\rm jet}\approx 300\,{\rm ms^{-1}}$, which is still too low for this type of outflows. Using the orbital timescale of the binary, $t_{\rm orb}\approx 800$\,yr, we get $v_{\rm jet}\approx 6\,{\rm kms^{-1}}$ -- i.e. at least one order of magnitude smaller than $v_{\rm jet}\approx 50\textrm{--}200\,{\rm kms^{-1}}$ we discussed above.

We conclude that any mechanism responsible for the launch of the jet in IRAS04158+2805 must act on a timescale shorter than the binary orbital timescale. One possibility could be a boost in the accretion rate when the binary passes through the pericentre of the orbit.
In order to be consistent with the typical velocities observed in proto-stellar jets -- assuming a typical $v_{\rm jet}\approx 100 ,{\rm kms^{-1}}$, the jet precession timescale should be of $\approx 50$ yr (i.e. $<10\%$ of the binary orbital timescale).

We caution that the available data are not sufficient to formulate a reliable hypothesis, or to exclude possible mechanisms launching the jet. Future dedicated studies are planned to address the issue.

\section{Summary and conclusions}\label{sec:conclusions}

In this paper, we presented unpublished observations complemented with additional data from the literature to develop a cohesive dynamical model of the system IRAS~04158+2805.
This system presents a resolved low mass binary in the centre of a large circumbinary disc. The circumbinary disc has a large lopsided cavity, with semi-major axis $a_{\rm cav}\sim 185$ au and eccentricity $e_{\rm cav}\sim 0.2$, in the ALMA band 7 continuum. Two marginally resolved circumstellar discs are also detected at the center (ALMA Band 4 and Band 7, see Figs. \ref{fig:contB4B7obs}, for more details see Sec. \ref{sec:obs}).
Bright CO $J=3\textrm{--}2$ line emission from the central region of the system suggest that some gas is present in the surroundings of the binary (Fig. \ref{fig:MmapsObs}). Kinematic data from CO emission appear to confirm the previous estimates of the binary mass from \citet{andrews2008} of $M_{\rm \star,tot}\sim 0.3\,{\rm M_\odot}$, but do not show signs of asymmetry due to the cavity eccentricity (see P--V diagram in Fig. \ref{fig:pvdiag}). Higher spectral/spatial resolution observations of CO line emission are required to assess the real nature of the eccentric geometry of the dust cavity.
A low contrast ratio azimuthal over-dense feature is present at the western edge of the cavity.

We present and analyse the astrometric data of the binary hosted in the system by fitting the observed arc of the orbit using \textsc{imorbel} \citep{pearce2015}. We further refine the estimates on the binary orbital parameters, by comparing the disc structures observed by ALMA with the theoretical predictions from binary-disc interaction theory \citep{pichardo2005,pichardo2008,miranda2015}. We also perform an estimate of the dust mass in the system based on the total flux detected in the circumbinary and circumstellar discs; we find a circumbinary dust mass $M_{\rm dust}\sim 1.5\times 10^{-4}\,{\rm M_\odot}$ which suggests a gas mass of $M_{\rm gas}\approx 15\,{\rm M_J}$, assuming a gas-to-dust ratio $M_{\rm gas}/M_{\rm dust}=100$.

The analysis of the binary astrometric data (Sec. \ref{sec:binorbastrom}) favours large values of binary eccentricity ($e>0.5$), a moderate binary-disc mutual inclination ($\theta\lesssim 30^\circ$) and semi-major axes $a<100$ au. Our dynamical considerations from binary disc interaction theory we provided in Sec. \ref{sec:binorbbindisc} further constrain the binary eccentricity value to $0.5< e < 0.7$, since values $e>0.7$ would imply circumstellar discs inconsistent with the observed ones --- $R_{{\rm cs},i}<3$ au. Using a similar approach (see Fig. \ref{fig:avse}), we constrain the binary semi-major axis to $a\approx 55$ au , while its inclination with respect to the disc orbital plane appears to exclude values $\theta> 30^\circ$.

Based on the inferred parameters, we perform three numerical simulations using the SPH code \textsc{phantom} \citep{price2018a} using $a=55$ au, $e=0.5$ and three different mutual binary-disc inclinations $\theta=\{0^\circ,\,30^\circ,\,60^\circ\}$ (see Fig. \ref{fig:snapshots}).
We perform Monte Carlo radiative transfer using \textsc{mcfost} \citep{pinte2006} on the snapshots from these three simulations after $\approx 40$ binary orbits and compare the results with observations (Fig. \ref{fig:continuum2}, \ref{fig:COmaps} and \ref{fig:cuts}), with the goal to directly compare the models with the observations.

Moderate inclinations ($\theta=0^\circ\textrm{--}30^\circ$) show a good agreement both in the morphology of the emission --- i.e., cavity and azimuthal over-density are well reproduced by our simulations --- and in the flux scale. The high inclination case ($\theta=60^\circ$) does not reproduce well the characteristic features of the system.

Our synthetic observations do not show a dust cavity eccentricity exceeding $e_{\rm cav}\sim 0.05$ in contrast with what observed. However, we note that the simulations snapshots of the cases $\theta=0^\circ\textrm{--}30^\circ$ do show some degree of lopsidedness in the dust cavity that suggests $e_{\rm cav}\sim 0.2$ from a measure of the difference between apocentre and pericentre (see Fig. \ref{fig:snapshots}), making the discrepancy due to a projection effect.

Overall, we conclude that the best agreement between our model and the observations is obtained for the case with binary-disc inclination $\theta=0^\circ$ and $\theta=30^\circ$. We can safely exclude the $\theta=60^\circ$ case, because both the reduced cavity size and the lack of azimuthal over-density make this case significantly different from the observations. Non-coplanarity between the binary and the circumbinary disc has been previously invoked to explain the morphology in a number of systems \citep{price2018,calcino2020,poblete2020,ballabio2021}. This highlights that misaligned geometries are not particularly exotic and is in agreement with the results by \citet{czekala2019}, who found that distribution of binary-disc inclination is relatively flat for long period binaries.

Theoretical predictions from binary-disc interaction theory provide a useful tool to infer the properties of binaries based on the morphology of the structures in the material surrounding them.

\section*{Acknowledgements}

We thank the anonymous referee for carefully reading the paper and for their constructive comments.

ER and CT are grateful to Linda Podio for insightful discussion about jets in YSO.

ER acknowledges financial support from the European Research Council (ERC) under the European Union's Horizon 2020 research and innovation programme (grant agreement No 681601). GD acknowledges support from NASA grant 80NSSC18K0442. CP acknowledges funding from the Australian Research Council via
FT170100040 and DP180104235. MV research was supported by an appointment to the NASA Postdoctoral Program at the NASA Jet Propulsion Laboratory, administered by Universities Space Research Association under contract with NASA.

This project has received funding from the European Union's Horizon 2020 research and innovation programme under the Marie Sk\l{}odowska-Curie grant agreement No 210021.
This project has received funding from the European Union's Horizon 2020 research and innovation programme under the Marie Sk\l{}odowska-Curie grant agreement No 823823 (DUSTBUSTERS).

The simulations performed for this paper used the DiRAC Data Intensive service at Leicester, operated by the University of Leicester IT Services, which forms part of the STFC DiRAC HPC Facility (www.dirac.ac.uk). The equipment was funded by BEIS capital funding via STFC capital grants ST/K000373/1 and ST/R002363/1 and STFC DiRAC Operations grant ST/R001014/1. DiRAC is part of the National e-Infrastructure.
Fig. \ref{fig:snapshots} was created using \textsc{splash} \citep{price07a}. All the other figures were created using \textsc{matplotlib} python library \citep{hunter2007}.

\section*{Data Availability Statement}

The SPH code \textsc{phantom}  is available at \url{https://github.com/danieljprice/phantom}. The observational data and the radiative transfer code \textsc{mcfost} are available upon request.



\bibliographystyle{mnras}
\bibliography{biblio} 




\appendix


\bsp	
\label{lastpage}
\end{document}